\pdfoutput=1

\PassOptionsToPackage{hyphens}{url}
\documentclass[format=sigconf,screen=true]{acmart}
\renewcommand\footnotetextcopyrightpermission[1]{} 
\usepackage{graphicx}
\usepackage{balance}

\usepackage{algorithm}
\usepackage[noend]{algpseudocode}
\usepackage{setspace}
\usepackage{listings}
\usepackage{multicol}
\usepackage{xspace}
\usepackage{etoolbox}
\usepackage{indentfirst}
\usepackage{enumitem}
\usepackage{boldline}
\usepackage{booktabs}
\newcommand{\ra}[1]{\renewcommand{\arraystretch}{#1}}

\newcommand{\minihead}[1]{{\vspace{.45em}\noindent\textbf{#1.} }}

\newcommand{\mh}{MinHash }
\newcommand{\minmax}{Min-Max hash}
\newcommand{\multiprobe}{multi-probe LSH }

\setcopyright{none}

\title{Locality-Sensitive Hashing for Earthquake Detection: \\A Case Study of Scaling Data-Driven Science}
\author{Kexin Rong\footnotemark[1], Clara E. Yoon\footnotemark[2], Karianne J. Bergen\footnotemark[3], 
 Hashem Elezabi\footnotemark[1], \\[-3mm]Peter Bailis\footnotemark[1], Philip Levis\footnotemark[1], Gregory C. Beroza\footnotemark[2]\\ [4mm]}\thanks{$\ast$ Department of Computer Science}\thanks{$\dagger$ Department of Geophysics}\thanks{$\ddagger$ Institute for Computational and Mathematical Engineering}
\affiliation{
    \institution{Stanford University}
}

\begin{document}

\begin{abstract}

In this work, we report on a novel application of Locality Sensitive Hashing (LSH) to seismic data at scale. Based on the high waveform similarity between reoccurring earthquakes, our application identifies potential earthquakes by searching for similar time series segments via LSH. However, a straightforward implementation of this LSH-enabled application has difficulty scaling beyond 3 months of continuous time series data measured at a single seismic station.
As a case study of a data-driven science workflow, we illustrate how domain knowledge can be incorporated into the workload to improve both the efficiency and result quality.
We describe several end-to-end optimizations of the analysis pipeline from pre-processing to post-processing, which allow the application to scale to time series data measured at multiple seismic stations. Our optimizations enable an over 100$\times$ speedup in the end-to-end analysis pipeline. This improved scalability enabled seismologists to perform seismic analysis on more than ten years of continuous time series data from over ten seismic stations, and has directly enabled the discovery of 597 new earthquakes near the Diablo Canyon nuclear power plant in California and 6123 new earthquakes in New Zealand.

\end{abstract}

\maketitle

\section{Introduction}

Locality Sensitive Hashing (LSH)~\cite{lsh} is a well studied computational primitive for efficient nearest neighbor search in high-dimensional spaces.
LSH hashes items into low-dimensional spaces such that similar items have a higher collision probability in the hash table.
Successful LSH applications include entity resolution~\cite{lsher}, genome sequence comparison~\cite{lshgenome}, text and image search~\cite{lshtextsearch,lshimagesearch}, near duplicate detection~\cite{neardup,googlededup}, and video identification~\cite{lshvideo}.

In this paper, we present an innovative use of LSH---and associated challenges at scale---in large-scale earthquake detection across seismic networks.
Earthquake detection is particularly interesting in both its abundance of raw data and scarcity of labeled examples:

First, seismic data is large. Earthquakes are monitored by seismic networks, which can contain thousands of seismometers that continuously measure ground motion and vibration.
For example, Southern California alone has over 500 seismic stations, each collecting continuous ground motion measurements at 100Hz. As a result, this network alone has collected over ten trillion ($10^{13}$) data points in the form of time series in the past decade~\cite{caltech}.

Second, despite large measurement volumes, only a small fraction of earthquake events are cataloged, or confirmed and hand-labeled by domain scientists.
As earthquake magnitude (i.e., size) decreases, the frequency of earthquake events increases exponentially. Worldwide, major earthquakes (magnitude 7+) occur approximately once a month, while magnitude 2.0 and smaller earthquakes can occur several thousand times a day. 
At low magnitudes, it is increasingly difficult to detect earthquake signals because earthquake energy approaches the noise floor, and conventional seismological analyses can fail to disambiguate between signal and noise.
Nevertheless, detecting these small earthquakes is important in uncovering unknown seismic sources~\cite{smalleq4,FASTlarge}, improving the understanding of earthquake mechanics~\cite{smalleq1,smalleq3}, and better predicting the occurrences of future events~\cite{smalleq2}.

\begin{figure}
    \centering
    \includegraphics[width=\linewidth]{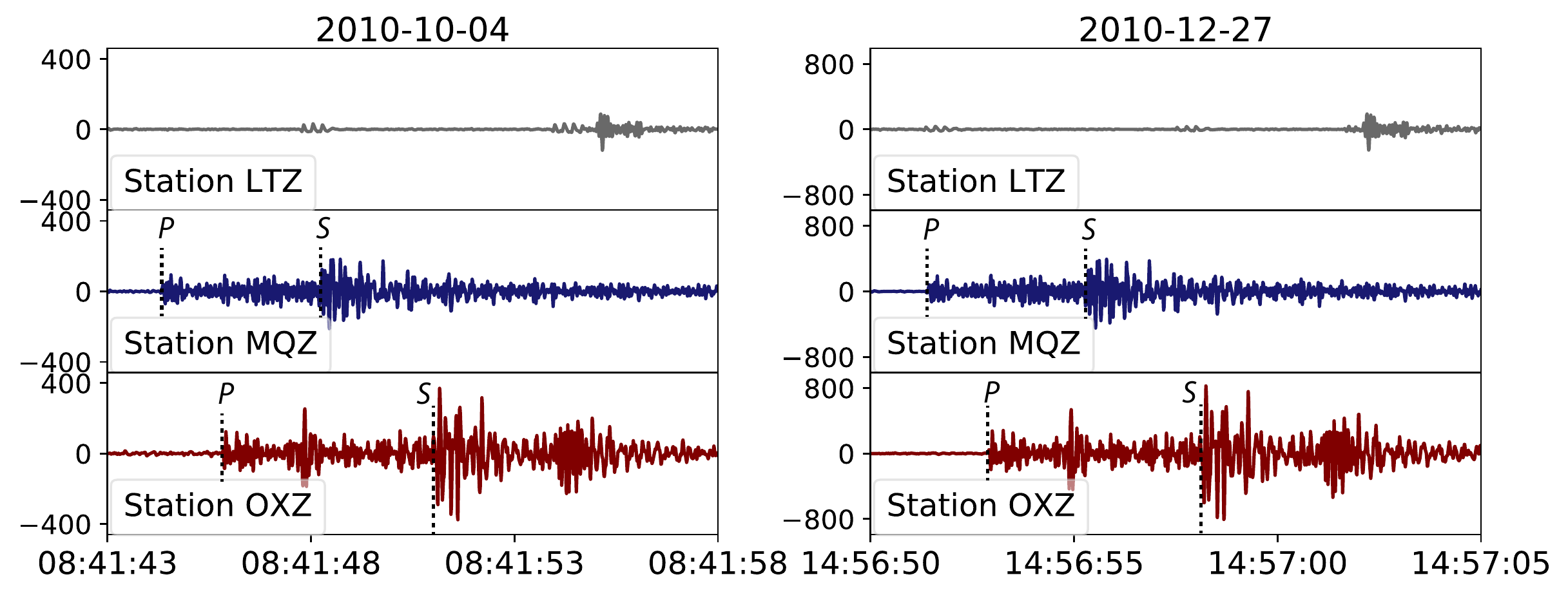}
    \caption{Example of near identical waveforms between occurrences of the same earthquake two months apart, observed at three seismic stations in New Zealand. The stations experience increased ground motions upon the arrivals of seismic waves (e.g., P and S waves). This paper scales LSH to over 30 billion data points and discovers 597 and 6123 new earthquakes near the Diablo Canyon nuclear power plant in California and in New Zealand, respectively. }
    \label{fig:similar_waveforms}
\vspace{-0.5em}
\end{figure}

To take advantage of the large volume of unlabeled raw measurement data, seismologists have developed an unsupervised, data-driven earthquake detection method, Fingerprint And Similarity Thresholding (FAST), based on waveform similarity~\cite{FAST}. Seismic sources repeatedly generate earthquakes over the course of days, months or even years, and these earthquakes show near identical waveforms when recorded at the same seismic station, regardless of the earthquake's magnitude~\cite{similarwaveform,similareq}.
Figure~\ref{fig:similar_waveforms} illustrates this phenomenon by depicting a pair of reoccurring earthquakes that are two months apart, observed at three seismic stations in New Zealand.
By applying LSH to identify similar waveforms from seismic data, seismologists were able to discover new, low-magnitude earthquakes without knowledge of prior earthquake events.

Despite early successes, seismologists had difficulty scaling their LSH-based analysis beyond 3-month of time series data ($7.95\times10^8$ data points) at a single seismic station~\cite{FASTlarge}. The FAST implementation faces severe scalability challenges. Contrary to what LSH theory suggests, the actual LSH runtime in FAST grows near quadratically with the input size due to correlations in the seismic signals: in an initial performance benchmark, the similarity search took 5 CPU-days to process 3 months of data, and, with a 5$\times$ increase in dataset size, LSH query time increased by 30$\times$.
In addition, station-specific repeated background noise leads to an overwhelming number of similar but non-earthquake time series matches, both crippling throughput and seismologists' ability to sift through the output, which can number in the hundreds of millions of events.
Ultimately, these scalability bottlenecks prevented seismologists from making use of the decades of data at their disposal.

In this paper, we show how systems, algorithms, and domain expertise can go hand-in-hand to deliver substantial scalability improvements for this seismological analysis.
Via algorithmic design, optimization using domain knowledge, and data engineering, we scale the FAST workload to years of continuous data at multiple stations.
In turn, this scalability has enabled new scientific discoveries, including previously unknown earthquakes near a nuclear reactor in San Luis Obispo, California, and in New Zealand.


Specifically, we build a scalable end-to-end earthquake detection pipeline comprised of three main steps.
First, the fingerprint extraction step encodes time-frequency features of the original time series into compact binary fingerprints that are more robust to small variations.
To address the bottleneck caused by repeating non-seismic signals, we apply domain-specific filters based on the frequency bands and the frequency of occurrences of earthquakes.
Second, the search step applies LSH on the binary fingerprints to identify all pairs of similar time series segments. We pinpoint high hash collision rates caused by physical correlations in the input data as a core culprit of LSH performance degradation and alleviate the impact of large buckets by increasing hash selectivity while keeping the detection threshold constant.
Third, the alignment step significantly reduces the size of detection results and confirms seismic behavior by performing spatiotemporal correlation with nearby seismic stations in the network~\cite{networkpaper}.
To scale this analysis, we leverage domain knowledge of the invariance of the time difference between a pair of earthquake events across all stations at which they are recorded.

In summary, as an innovative systems and applications paper, this work makes several contributions:
\begin{itemize}[noitemsep,topsep=.3em]
 \setlength\itemsep{0.1em}
    \item We report on a new application of LSH in seismology as well as a complete end-to-end data science pipeline, including non-trivial pre-processing and post-processing, that scales to a decade of continuous time series for earthquake detection.
    \item We present a case study for using domain knowledge to improve the accuracy and efficiency of the pipeline. We illustrate how applying seismological domain knowledge in each component of the pipeline is critical to scalability.
    \item We demonstrate that our optimizations enable a cumulative two order-of-magnitude speedup in the end-to-end detection pipeline. These quantitative improvements enable qualitative discoveries: we discovered 597 new earthquakes near the Diablo Canyon nuclear power plant in California and 6123 new earthquakes in New Zealand, allowing seismologists to determine the size and shape of nearby fault structures.
\end{itemize}
Beyond these contributions to a database audience, our solution is an open source tool, available for use by the broader scientific community. We have already run workshops for seismologists at Stanford~\cite{fastgithub} and believe that the pipeline can not only facilitate targeted seismic analysis but also contribute to the label generation for supervised methods in seismic data~\cite{convquake}.

The rest of the paper proceeds as follows. We review background information about earthquake detection in Section 2 and discuss additional related work in Section 3. We give a brief overview of the end-to-end detection pipeline and key technical challenges in Section 4. Sections 5, 6 and 7 present details as well as optimizations in the fingerprint extraction, similarity search and the spatiotemporal alignment steps of the pipeline. We perform a detailed evaluation on both the quantitative performance improvements of our optimizations as well as qualitative results of new seismic findings in Section 8. In Section 9, we reflect on lessons learned and conclude.

\section{Background}
\label{sec:bg}

With the deployment of denser and increasingly sensitive sensor arrays, seismology is experiencing a rapid growth of high-resolution data~\cite{array}. Seismic networks with up to thousands of sensors have been recording years of continuous seismic data streams, typically at 100Hz frequencies. The rising data volume has fueled strong interest in the seismology community to develop and apply scalable data-driven algorithms that improve the monitoring and prediction of earthquake events~\cite{dmreview,PCA,myshake}.

In this work, we focus on the problem of detecting new, low-magnitude earthquakes from historical seismic data. Earthquakes, which are primarily caused by the rupture of geological faults, radiate energy that travels through the Earth in the form of seismic waves. Seismic waves induce ground motion that is recorded by seismometers. Modern seismometers typically include 3 components that measure simultaneous ground motion along the north-south, east-west, and vertical axes. Ground motions along each of these three axes are recorded as a separate \emph{channel} of time series data.

Channels capture complementary signals for different seismic waves, such as the P-wave and the S-wave. The P-waves travel along the direction of propagation, like sound, while the S-waves travel perpendicular to the direction of propagation, like ocean waves. The vertical channel, therefore, better captures the up and down motions caused by the P-waves while the horizontal channels better capture the side to side motions caused by the S-waves. P-waves travel the fastest and are the first to arrive at seismic stations, followed by the slower but usually larger amplitude S-waves. Hence, the P-wave and S-wave of an earthquake typically register as two ``big wiggles" on the ground motion measurements (Figure~\ref{fig:similar_waveforms}). These impulsive arrivals of seismic waves are example characteristics of earthquakes that seismologists look for in the data.

While it is easy for human eyes to identify large earthquakes on a single channel, accurately detecting small earthquakes usually requires looking at data from multiple channels or stations. These low-magnitude earthquakes pose challenges for conventional methods for detection, which we outline below. Traditional energy-based earthquake detectors such as a short-term average (STA)/long-term average (LTA) identify earthquake events by their impulsive, high signal-to-noise P-wave and S-wave arrivals. However, these detectors are prone to high false positive and false negative rates at low magnitudes, especially with noisy backgrounds~\cite{staltalow}. Template matching, or the waveform cross-correlation with template waveforms of known earthquakes, has proven more effective for detecting known seismic signals in noisy data~\cite{template1,template2}. However, the method relies on template waveforms of prior events and is not suitable for discovering events from unknown sources.

As a result, almost all earthquakes greater than magnitude 5 are detected~\cite{mag5}. In comparison, an estimated 1.5 million earthquakes with magnitude between 2 and 5 are not detected by conventional means, and 1.3 million of these are between magnitude 2 and 2.9. The estimate is based on the magnitude frequency distribution of earthquakes~\cite{grlaw}.  We are interested in detecting these low-magnitude earthquakes missing from public earthquake catalogs to better understand earthquake mechanics and sources, which inform seismic hazard estimates and prediction~\cite{smalleq1,smalleq2,smalleq3,smalleq4}.

The earthquake detection pipeline we study in the paper is an unsupervised and data-driven approach that does not rely on supervised (i.e., labeled) examples of prior earthquake events, and is designed to complement existing, supervised detection methods. As in template matching, the method we optimize takes advantage of the high similarity between waveforms generated by reoccurring earthquakes. However, instead of relying on waveform templates from only known events, the pipeline leverages the recurring nature of seismic activities to detect similar waveforms in time and across stations. To do so, the pipeline performs an all-pair time series similarity search, treating each segment of the input waveform data as a ``template" for potential earthquakes. The proposed approach can not detect an earthquake that occurs only once and is not similar enough to any other earthquakes in the input data. Therefore, to improve detection recall, it is critical to be able to scale the analysis to input data with a longer duration (e.g., years instead of weeks or months).

\section{Related Work}
In this section, we address related work in earthquake detection, LSH-based applications and time series similarity search.

\minihead{Earthquake Detection} The original FAST work appeared in the seismology community, and has proven a useful tool in scientific discovery~\cite{FAST, FASTlarge}. In this paper, we present FAST to a database audience for the first time, and report on both the pipeline composition and optimization from a computational perspective. The results presented in this paper are the result of over a year of collaboration between our database research group and the Stanford earthquake seismology research group. The optimizations we present in this paper and the resulting scalability results of the optimized pipeline have not previously been published. We believe this represents a useful and innovative application of LSH to a real domain science tool that will be of interest to both the database community and researchers of LSH and time-series analytics.

The problem of earthquake detection is decades old~\cite{oldtextbook}, and many classic techniques---many of which are in use today---were developed for an era in which humans manually inspected seismographs for readings~\cite{stalta1, stalta2}. With the rise of machine learning and large-scale data analytics, there has been increasing interest in further automating these techniques. While FAST is optimized to find many small-scale earthquakes, alternative approaches in the seismology community utilize template matching~\cite{template1, template2}, social media~\cite{twitter}, and machine learning techniques~\cite{nn, svm} to detect earthquakes. Most recently, with sufficient training data, supervised approaches have shown promising results of being able to detect non-repeating earthquake events~\cite{convquake}. In contrast, our LSH-based detection method does not rely on labeled earthquake events and detects reoccurring earthquake events. In the evaluation, we compare against two supervised methods~\cite{weasel,convquake} and show that our unsupervised pipeline is able to detect qualitatively different events from existing earthquake catalogs. 

\minihead{Locality Sensitive Hashing} In this work, we perform a detailed case study of the practical challenges and the domain-specific solutions of applying LSH to the field of seismology. We do not contribute to the advance of the state-of-the-art LSH algorithms; instead, we show that classic LSH techniques, combined with domain-specific optimizations, can lead to scientific discoveries when applied at scale. Existing work shows that LSH performance is sensitive to key parameters such as the number of hash functions~\cite{lshtextsearch, lshmodel}; we provide supporting evidence and analysis on the performance implication of LSH parameters in our application domain. In addition to the core LSH techniques, we also present nontrivial preprocessing and postprocessing steps that enable an end-to-end detection pipeline, including spatiotemporal alignment of LSH matches.

Our work targets CPU workloads, complementing existing efforts that speed up similarity search on GPUs~\cite{searchgpu}. To preserve the integrity of the established science pipeline, we focus on optimizing the existing \mh based LSH rather than replacing it with potentially more efficient LSH variants such as LSH forest~\cite{lshforest} and \multiprobe~\cite{multiprobe}. While we share observations with prior work that parallelizes and distributes a different LSH family~\cite{twitterlsh}, we present the unique challenges and opportunities of optimizing MinHash LSH in our application domain. We provide performance benchmarks against alternative similarity search algorithms in the evaluation, such as set similarity joins~\cite{setsimilarity} and an alternative LSH library based on recent theoretical advances in LSH for cosine similarity~\cite{falconn}. We believe the resulting experience report, as well as our open source implementation, will be valuable to researchers developing LSH techniques in the future.

\minihead{Time Series Analytics} Time series analytics is a core topic in large-scale data analytics and data mining~\cite{clusteringsurvey, tsbench, classificationsurvey}. In our application, we utilize time series similarity search as a core workhorse for earthquake detection. There are a number of distance metrics for time series~\cite{distancesurvey}, including Euclidean distance and its variants~\cite{lp}, Dynamic Time Warping~\cite{trillion}, and edit distance~\cite{lcss}. However, our input time series from seismic sensors is high frequency (e.g. 100Hz) and often noisy. Therefore, small time-shifts, outliers and scaling can result in large changes in time-domain metrics~\cite{haar}. Instead, we encode time-frequency features of the input time series into binary vectors and focus on the Jaccard similarity between the binary feature vectors. This feature extraction procedure is an adaptation of the Waveprint algorithm~\cite{waveprint} initially designed for audio data; the key modification made for seismic data was to focus on frequency features that are the most discriminative from background noise, such that the average similarity between non-seismic signals is reduced~\cite{FASTFingerPrint}. An alternative binary representation models time series as points on a grid, and uses the non-empty grid cells as a set representation of the time series~\cite{setbased}. However, this representation does not take advantage of the physical properties distinguishing background from seismic signals.

\begin{figure*}[h!]
    \centering
    \includegraphics[width=\linewidth]{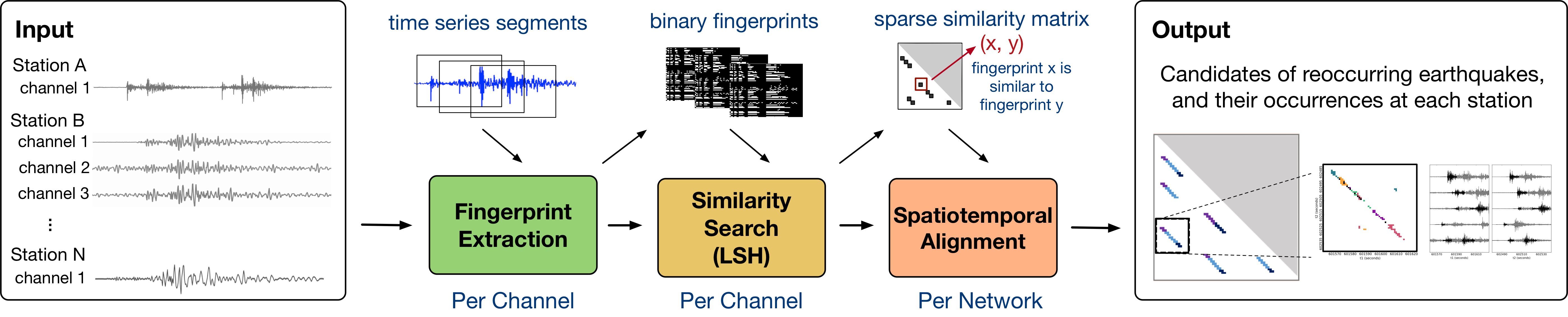}
    \vspace{-1em}
    \caption{The three steps of the end-to-end earthquake detection pipeline: fingerprinting transforms time series into binary vectors (Section~\ref{sec:fp}); similarity search identifies pairs of similar binary vectors (Section~\ref{sec:search}); alignment aggregates and reduces false positives in results (Section~\ref{sec:network}). }
    \label{fig:pipeline}
\end{figure*}

\section{Pipeline Overview}
\label{sec:overview}

In this section, we provide an overview of the three main steps of our end-to-end detection pipeline. We elaborate on each step---and our associated optimizations---in later sections, referenced inline.

The input of the detection pipeline consists of continuous ground motion measurements in the form of time series, collected from multiple stations in the seismic network. The output is a list of potential earthquakes, specified in the form of timestamps when the seismic wave arrives at each station. From there, seismologists can compare with public earthquake catalogs to identify new events, and visually inspect the measurements to confirm seismic findings.

Figure~\ref{fig:pipeline} illustrates the three major components of the end-to-end detection pipeline: fingerprint extraction, similarity search, and spatiotemporal alignment. For each input time series, or continuous ground motion measurements from a seismic channel, the algorithm slices the input into short windows of overlapping time series segments and encodes time-frequency features of each window into a binary fingerprint; the similarity of the fingerprints resembles that of the original waveforms (Section~\ref{sec:fp}). The algorithm then performs an all pairs similarity search via LSH on the binary fingerprints and identifies pairs of highly similar fingerprints (Section~\ref{sec:search}). Finally, like a traditional associator that maps earthquake detections at each station to a consistent seismic source, in the spatiotemporal alignment stage, the algorithm combines, filters and clusters the outputs from all seismic channels to generate a list of candidate earthquake detections with high confidence (Section~\ref{sec:network}).

A na\"ive implementation of the pipeline imposes several scalability challenges. For example, we observed LSH performance degradation in our application caused by the non-uniformity and correlation in the binary fingerprints; the correlations induce undesired LSH hash collisions, which significantly increase the number of lookups per similarity search query (Section~\ref{sec:searchparam}). In addition, the similarity search does not distinguish seismic from non-seismic signals. In the presence of repeating background signals, similar noise waveforms could outnumber similar earthquake waveforms, leading to more than an order of magnitude slow down in runtime and increase in output size (Section~\ref{sec:noise}). As the input time series and the output of the similarity search becomes larger, the pipeline must adapt to data sizes that are too large to fit into main memory (Section~\ref{sec:searchpart},~\ref{sec:networkimp}).

In this paper, we focus on single-machine, main-memory execution on commodity servers with multicore processors. We parallelize the pipeline within a given server but otherwise do not distribute the computation to multiple servers. In principle, the parallelization efforts extend to distributed execution. However, given the poor quadratic scalability of the unoptimized pipeline, distribution alone would not have been a viable option for scaling to desired data volume. As a result of the optimizations described in this paper, we are able to scale to a decade of data on a single node without requiring distribution. However, we view distributed execution as a valuable extension for future work.

In the remaining sections of this paper, we describe the design decisions as well as performance optimizations for each pipeline component. Most of our optimizations focus on the all pairs similarity search, where the initial implementation exhibited near quadratic growth in runtime with the input size. We show in the evaluation that, these optimizations enable speedups of more than two orders of magnitude in the end-to-end pipeline.

\section{Fingerprint Extraction}
\label{sec:fp}
In this section, we describe the fingerprint extraction step that encodes time-frequency features of the input time series into compact binary vectors for similarity search. We begin with an overview of the fingerprinting algorithm~\cite{FASTFingerPrint} and the benefits of using fingerprints in place of the time series (Section~\ref{sec:fpoverview}). We then describe a new optimization that parallelizes and accelerates the fingerprinting generation via sampling (Section~\ref{sec:mad}).

\subsection{Fingerprint Overview}
\label{sec:fpoverview}

Inspired by the success of feature extraction techniques for indexing audio snippets~\cite{FASTFingerPrint}, fingerprint extraction step transforms continuous time series data into compact binary vectors (fingerprints) for similarity search. Each fingerprint encodes representative time-frequency features of the time series. The Jaccard similarity of two fingerprints, defined as the size of the intersection of the non-zero entries divided by the size of the union, preserves the waveform similarity of the corresponding time series segments. Compared to directly computing similarity on the time series, fingerprinting introduces frequency-domain features into the detection and provides additional robustness against translation and small variations~\cite{FASTFingerPrint}.

\begin{figure}
    \centering
    \includegraphics[width=0.9\linewidth]{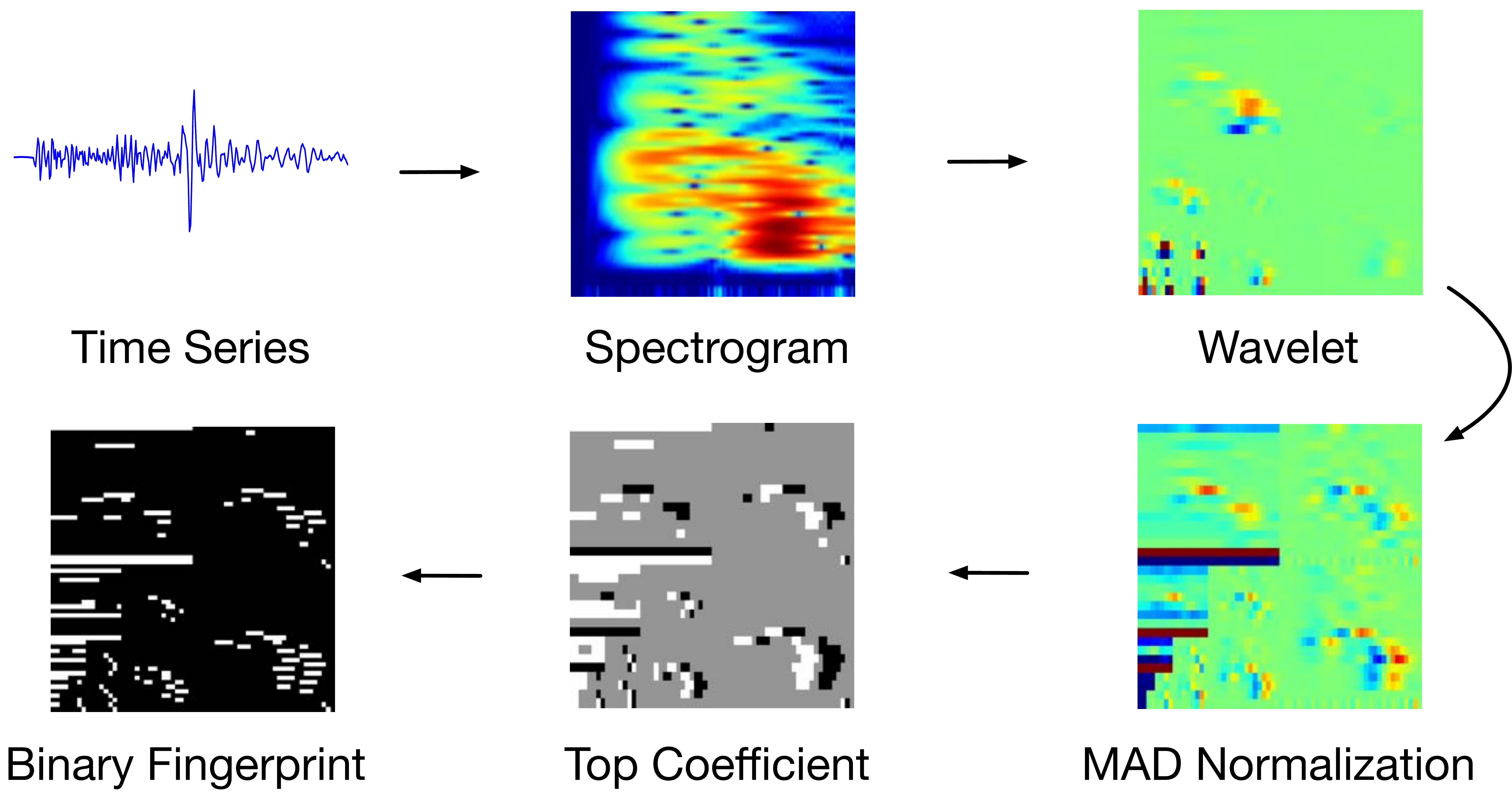}
        \caption{The fingerprinting algorithm encodes time-frequency features of the original time series into compact binary vectors.}
    \label{fig:fingerprint}
\end{figure}

Figure~\ref{fig:fingerprint} illustrates the individual steps of fingerprinting:
\begin{enumerate}[topsep=2pt]
    \item \textbf{Spectrogram} Compute the spectrogram, a time-frequency representation, of the time series. Slice the spectrogram into short overlapping segments using a sliding window and smooth by downsampling each segment into a spectral image of fixed dimensions.
    \item \textbf{Wavelet Transform} Compute two-dimensional discrete Haar wavelet transform on each spectral image. The wavelet coefficients are a lossy compression of the spectral images.
    \item \textbf{Normalization} Normalize each wavelet coefficient by its median and the median absolute deviation (MAD) on the full, background dominated dataset. 
    \item \textbf{Top coefficient} Extract the top K most anomalous wavelet coefficients, or the largest coefficients after MAD normalization, from each spectral image. 
    By selecting the most anomalous coefficients, we focus only on coefficients that are most distinct from coefficients that characterize noise, which empirically leads to better detection results.
    \item \textbf{Binarize} Binarize the signs and positions of the top wavelet coefficients. We encode the sign of each normalized coefficient using 2 bits:  $-1 \to$ 01, 0 $\to$ 00, 1 $\to$ 10.
\end{enumerate}

\subsection{Optimization: MAD via sampling}
\label{sec:mad}
The fingerprint extraction is implemented via scientific modules such as \texttt{scipy}, \texttt{numpy} and \texttt{PyWavelets} in Python. While its runtime grows linearly with input size, fingerprinting ten years of time series data can take several days on a single core. 

In the unoptimized procedure, normalizing the wavelet coefficients requires two full passes over the data. The first pass calculates the median and the MAD\footnote{For $X = \{x_1, x_2, ..., x_n\}$, the MAD is defined as the median of the absolute deviations from the median: $MAD = median(|x_i - median(X)|)$} for each wavelet coefficient over the whole population, and the second pass normalizes the wavelet representation of each fingerprint accordingly. Given the median and MAD for each wavelet coefficient, the input time series can be partitioned and normalized in parallel. Therefore, the computation of the median and MAD remains the runtime bottleneck.

We accelerate the computation by approximating the true median and MAD with statistics calculated from a small random sample of the input data. The confidence interval for MAD with a sample size of $n$ shrinks with $n^{1/2}$~\cite{madci}. We empirically find that, on one month of input time series data, sampling provides an order of magnitude speedup with almost no loss in accuracy. For input time series of longer duration (e.g. over a year), sampling 1\% or less of the input can suffice. We further investigate the trade-off between speed and accuracy under different sampling rates in the evaluation (Section~\ref{eval:params}, Appendix~\ref{appendix:eval}).
\section{LSH-based Similarity Search}
\label{sec:search}
In this section, we present the time series similar search step based on LSH. We start with a description of the algorithm and the baseline implementation (Section~\ref{sec:searchimpl}), upon which we build the optimizations. Our contributions include: an optimized hash signature generation procedure (Section~\ref{sec:hashgen}), an empirical analysis of the impact of hash collisions and LSH parameters on query performance (Section~\ref{sec:searchparam}),  partition and parallelization of LSH that reduce the runtime and memory usage (Section~\ref{sec:searchpart}), and finally, two domain-specific filters that improve both the performance and detection quality of the search (Section~\ref{sec:noise}).

\subsection{Similarity Search Overview}
\label{sec:searchimpl}

Reoccurring earthquakes originated from nearby seismic sources appear as near-identical waveforms at the same seismic station. Given continuous ground motion measurements from a seismic station, our pipeline identifies similar time series segments from the input as candidates for reoccurring earthquake events. 

\begin{figure}[t]
\centering
\includegraphics[width=\linewidth]{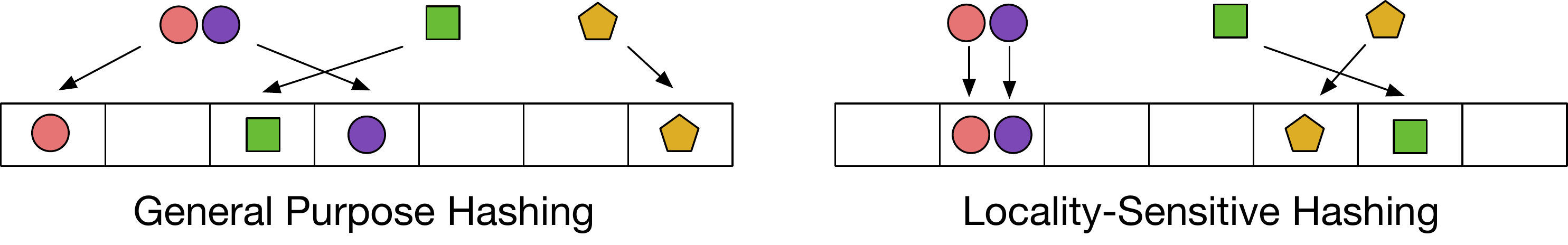}
\vspace{-1.5em}
\caption{Locality-sensitive hashing hashes similar items to the same hash ``bucket" with high probability.  }
\label{fig:lsh}
\end{figure}

Concretely, we perform an approximate similarity search via \mh LSH on the binary fingerprints to identify all pairs of fingerprints whose Jaccard similarity exceeds a predefined threshold~\cite{minhash}. \mh LSH performs a random projection of high-dimensional data into lower dimensional space, hashing similar items to the same hash table ``bucket" with high probability (Figure~\ref{fig:lsh}). Instead of performing a na\"ive pairwise comparisons between all fingerprints, LSH limits the comparisons to fingerprints sharing the same hash bucket, significantly reducing the computation. The ratio of the average number of comparisons per query to the size of the dataset, or \emph{selectivity}, is a machine-independent proxy for query efficiency~\cite{lshmodel}. 

\minihead{Hash signature generation} The \mh of a fingerprint is the first non-zero element of the fingerprint under a given random permutation of its elements. The permutation is defined by a hash function mapping fingerprint elements to random indices. Let $p$ denote the collision probability of a hash signature generated with a single hash function. By increasing the number of hash functions $k$, the collision probability of the hash signature decreases to $p^k$~\cite{mmds}. 

\minihead{Hash table construction} Each hash table stores an independent mapping of fingerprints to hash buckets. The tables are initialized by mapping hash signatures to a list of fingerprints that share the same signature. Empirically, we find that using $t=100$ hash tables suffices for our application, and there is little gain in further increasing the number of hash tables.

\minihead{Search} The search queries the hash tables for each fingerprint's near neighbor candidates, or other fingerprints that share the query fingerprint's hash buckets. We keep track of the number of times the query fingerprint and candidates have matching hash signature in the hash tables, and output candidates with matches above a predefined threshold. The number of matches is also used as a proxy for the confidence of the similarity in the final step of the pipeline.

\subsection{Optimization: Hash signature generation}
\label{sec:hashgen}

In this subsection, we present both memory access pattern and algorithmic improvements to speed up the generation of hash signatures. We show that, together, the optimizations lead to an over 3$\times$ improvement in hash generation time (Section~\ref{sec:e2e}).

Similar to observations made for SimHash (a different hash family for angular distances)~\cite{twitterlsh}, a na\"ive implementation of the MinHash generation can suffer from poor memory locality due to the sparsity of input data. SimHash functions are evaluated as a dot product between the input and hash mapping vectors, while \mh functions are evaluated as a minimum of hash mappings corresponding to non-zero elements of the input. For sparse input, both functions access scattered, non-contiguous elements in the hash mapping vector, causing an increase in cache misses. We improve the memory access pattern by blocking the access to the hash mappings. We use dimensions of the fingerprint, rather than hash functions, as the main loop for each fingerprint. As a result, the lookups for each non-zero element in the fingerprint are blocked into rows in the hash mapping array. For our application, this loop order has the additional advantage of exploiting the high overlap (e.g. over 60\% in one example) between neighboring fingerprints. The overlap means that previously accessed elements in hash mappings are likely to get reused while in cache, further improving the memory locality. 

In addition, we speed up the hash signature generation by replacing \mh with \minmax. MinHash only keeps the minimum value for each hash mapping, while \minmax keeps both the min and the max. Therefore, to generate hash signatures with similar collision probability, Min-Max hash reduces the number of required hash functions to half. Previous work showed the Min-Max hash is an unbiased estimator of pairwise Jaccard similarity, and achieves similar and sometimes smaller mean squared error (MSE) in estimating pairwise Jaccard similarity in practice~\cite{minmaxhash}. We include pseudocode for the optimized hash signature calculation in Appendix~\ref{appendix:hash} of this report.

\begin{figure}[t!]
    \centering
    \includegraphics[width=0.9\linewidth]{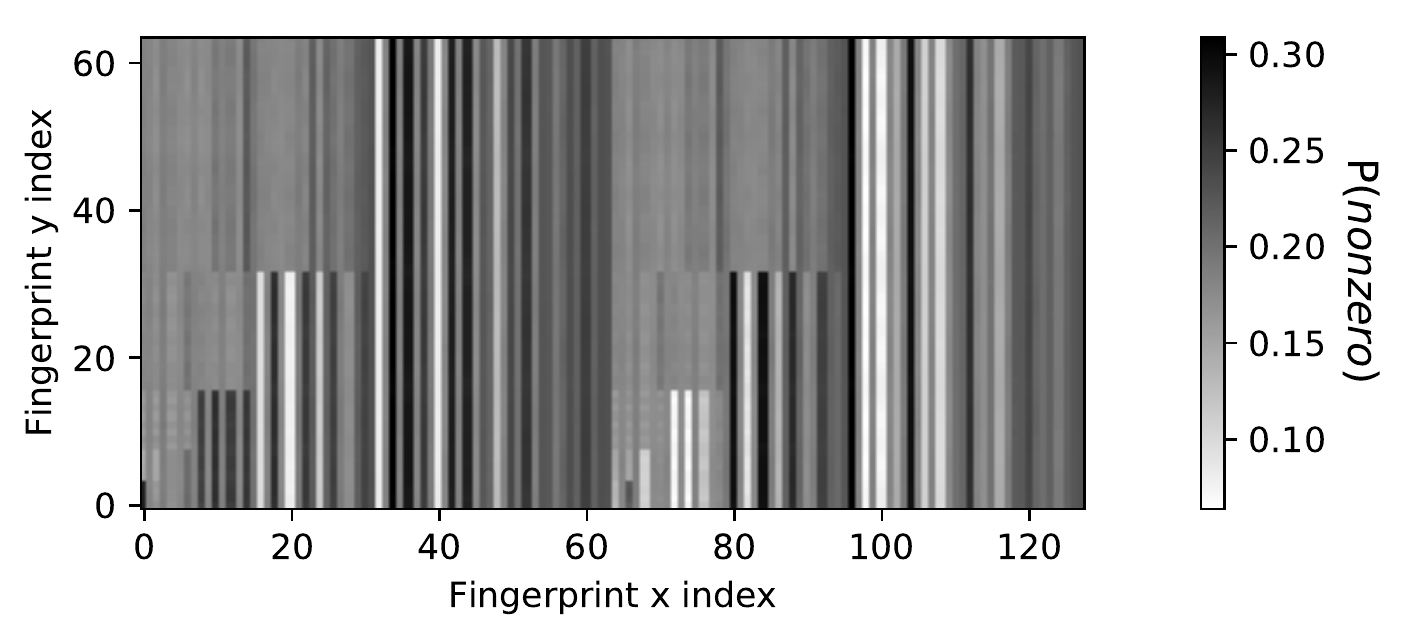}
    \vspace{-1em}
    \caption{Probability that each element in the fingerprint is equal to 1, averaged over 15.7M fingerprints, each of dimension 8192, generated from a year of time series data. The heatmap shows that some elements of the fingerprint are much more likely to be non-zero compared to others. }
    \label{fig:imbalance}
\end{figure}

\subsection{Optimization: Alleviating hash collisions}
\label{sec:searchparam}

Perhaps surprisingly, our initial LSH implementation demonstrated poor scaling with the input size: with a 5$\times$ increase in input, the runtime increases by 30$\times$. In this subsection, we analyze the cause of LSH performance degradation and the performance implications of core LSH parameters in our application.

\minihead{Cause of hash collisions} Poor distribution of hash signatures can lead to large LSH hash buckets or high query \emph{selectivity}, significantly degrading the performance of the similarity search~\cite{lshforest, lshskew}. For example, in the extreme case when all fingerprints are hashed into a single bucket, the \emph{selectivity} equals 1 and the LSH performance is equivalent to that of the exhaustive $O(n^2)$ search.

Our input fingerprints encode physical properties of the waveform data. As a result, the probability that each element in the fingerprint is non-zero is highly non-uniform (Figure~\ref{fig:imbalance}). Moreover, fingerprint elements are not necessarily independent, meaning that certain fingerprint elements are likely to co-occur: given an element $a_i$ is non-zero, the element $a_j$ has a much higher probability of being non-zero ($\mathbb{P}[a_i = 1, a_j = 1] > \mathbb{P}[a_i = 1] \times  \mathbb{P}[a_j = 1]$).

This correlation has a direct impact on the collision probability of \mh signatures. For example, if a hash signature contains $k$ independent \mh of a fingerprint and two of the non-zero elements responsible for the \mh are dependent, then the signature has effectively similar collision probability as the signature with only $k-1$ \mh. In other words, more fingerprints are likely to be hashed to the same bucket under this signature. For fingerprints shown in Figure~\ref{fig:imbalance}, the largest 0.1\% of the hash buckets contain an average of 32.9\% of the total fingerprints for hash tables constructed with $6$ hash functions.

\minihead{Performance impact of LSH parameters} The precision and recall of the LSH can be tuned via two key parameters: the number of hash functions $k$ and the number of hash table matches $m$. Intuitively, using $k$ hash functions is equivalent to requiring two fingerprints agree at $k$ randomly selected non-zero positions. Therefore, the larger the number of hash functions, the lower the probability of collision. To improve recall, we increase the number of independent permutations to make sure that similar fingerprints can land in the same hash bucket with high probability.

\begin{figure}[t!]
    \centering
    \includegraphics[width=0.9\linewidth]{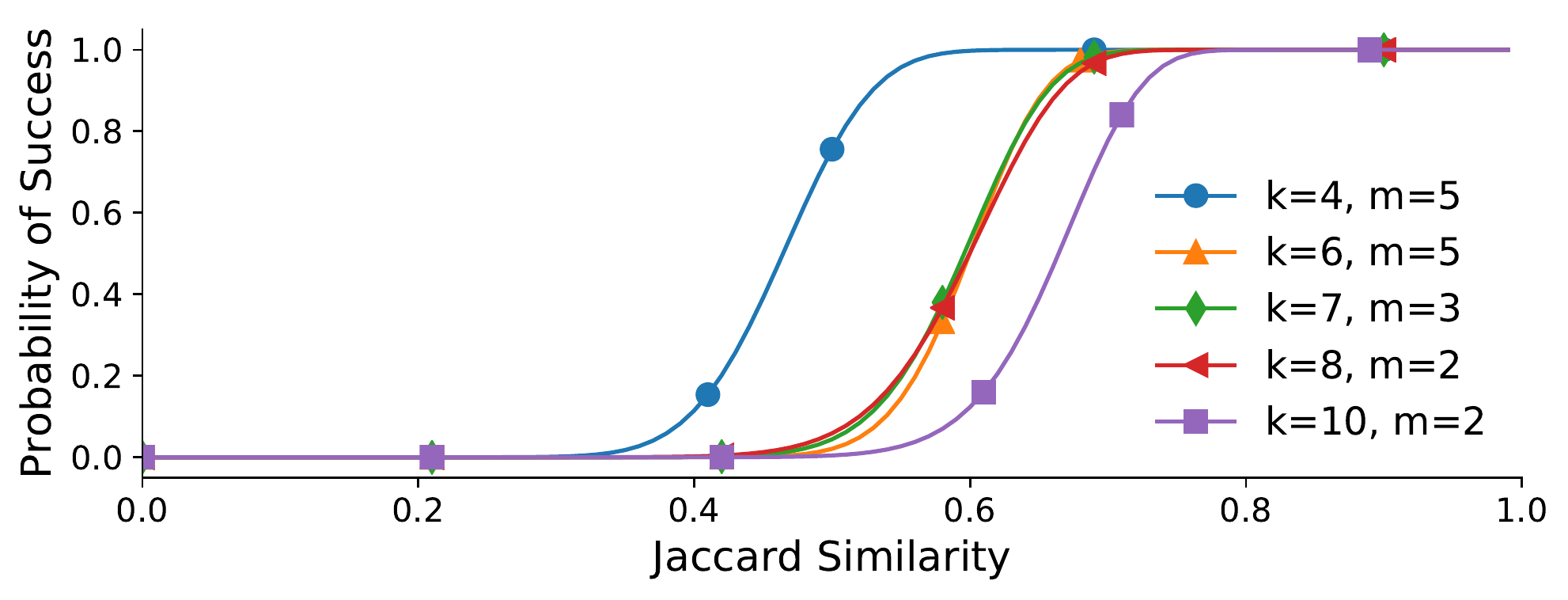}
    \vspace{-1em}
    \caption{Theoretical probability of a successful search versus Jaccard similarity between fingerprints ($k$: number of hash functions, $m$: number of matches). Different LSH parameter settings can have near identical detection probability with vastly different runtime.}
    \label{fig:prob}
\end{figure}

Formally, given two fingerprints with Jaccard similarity $s$, the probability that with $k$ hash functions, the fingerprints are hashed to the same bucket at least $m$ times out of $t=100$ hash tables is:
\[\mathbb{P}[s] = 1 - \sum_{i = 0} ^{m - 1} [\binom{t}{i}(1 - s^{k})^{t-i}(s^{k})^i].\]
The probability of detection success as a function of Jaccard similarity has the form of an S-curve (Figure~\ref{fig:prob}). The S-curve shifts to the right with the increase in the number of hash functions $k$ or the number of matches $m$, increasing the Jaccard similarity threshold for LSH. Figure~\ref{fig:prob} illustrates a near-identical probability of success curve under different parameter settings.

Due to the presence of correlations in the input data, LSH parameters with the same theoretically success probability can have vastly different runtime in practice. Specifically, as the number of hash functions increases, the expected average size of hash buckets decreases, which can lead to an order of magnitude speed up in the similarity search for seismic data in practice. However, to keep the success probability curve constant with increased hash functions, the number of matches needs to be lowered, which increases the probability of spurious matches. These spurious matches can be suppressed by scaling up the number of total hash tables, at the cost of larger memory usage. We further investigate the performance impact of LSH parameters in the evaluation.

\subsection{Optimization: Partitioning}
\label{sec:searchpart}
In this subsection, we describe the partition and parallelization of the LSH that further reduce its runtime and memory footprint. 

\minihead{Partition} Using a 1-second lag for adjacent fingerprints results in around 300M total fingerprints for 10 years of time series data. Given a hash signature of 64 bits and 100 total hash tables, the total size of hash signatures is approximately 250 GB. To avoid expensive disk I/O, we also want to keep all hash tables in memory for lookups. Taken together, this requires several hundred gigabytes of memory, which can exceed available main memory.

To scale to larger input data on a single node with the existing LSH implementation, we perform similarity search in partitions. We evenly partition the fingerprints and populate the hash tables with one partition at a time, while still keeping the lookup table of fingerprints to hash signatures in memory. 
During query, we output matches between fingerprints in the current partition (or in the hash tables) with all other fingerprints and subsequently repeat this process for each partition. The partitioned search yields identical results to the original search, with the benefit that only a subset of the fingerprints are stored in the hash tables in memory. We can partition the lookup table of hash signatures similarly to further reduce memory. We illustrate the performance and memory trade-offs under different numbers of partitions in Section~\ref{eval:params}.

The idea of populating the hash table with a subset of the input could also be favorable for performing a small number of nearest neighbor queries on a large dataset, e.g., a thousand queries on a million items. There are two ways to execute the queries. We can hash the full dataset and then perform a thousand queries to retrieve near neighbor candidates in each query item's hash buckets; alternatively, we can hash only the query items and for every other item in the dataset, check whether it is mapped to an existing bucket in the table. While the two methods yield identical query results, the latter could be $8.6\times$ faster since the cost of initializing the hash table dominates that of the search. 

It is possible to further improve LSH performance and memory usage with the more space efficient variants such as multi-probe LSH~\cite{multiprobe}. However, given that the alignment step uses the number of hash buckets shared between fingerprints as a proxy for similarity, and that switching to a multi-probe implementation would alter this similarity measure, we preserve the original LSH implementation for backwards compatibility with FAST. We compare against alternative LSH implementations and demonstrate the potential benefits of adopting \multiprobe in the evaluation (Section~\ref{sec:falconn}).

\minihead{Parallelization}
Once the hash mappings are generated, we can easily partition the input fingerprints and generate the hash signatures in parallel. Similarly, the query procedure can be parallelized by running nearest neighbor queries for different fingerprints and outputting results to files in parallel. We show in Section~\ref{eval:params} that the total hash signature generation time and similarity search time reduces near linearly with the number of processes.

\subsection{Optimization: Domain-specific filters}
\label{sec:noise}
\begin{figure}
    \centering
    \includegraphics[width=\linewidth]{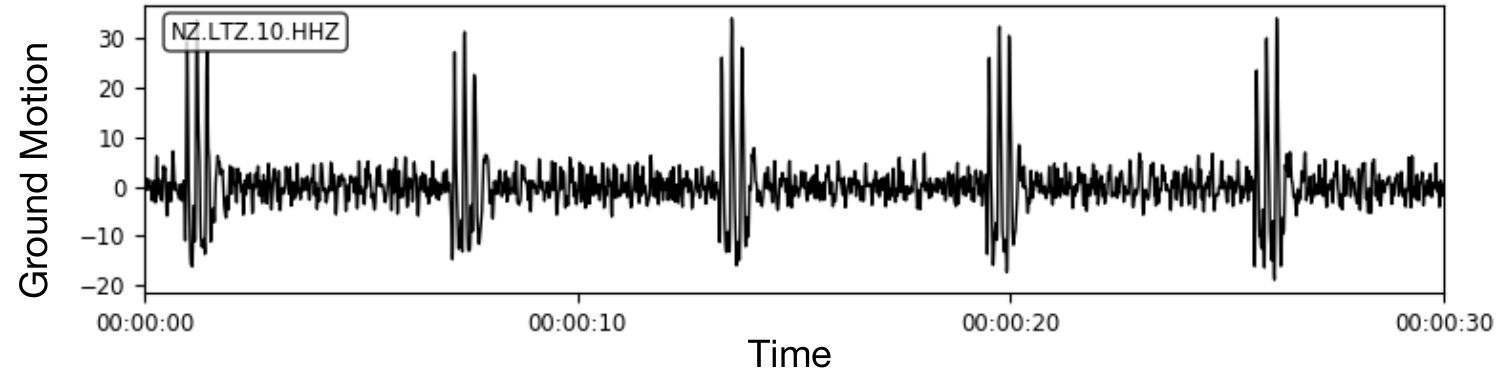}
    \vspace{-1.5em}
    \caption{The short, three-spike pattern is an example of similar and repeating background signals not due to seismic activity. These repeating noise patterns cause scalability challenges for LSH.}
    \label{fig:ltz_noise}
\end{figure}
Like many other sensor measurements, seismometer readings can be noisy. In this subsection, we address a practical challenge of the detection pipeline, where similar non-seismic signals dominate seismic findings in runtime and detection results. We show that by leveraging domain knowledge, we can greatly increase both the efficiency and the quality of the detection.

\minihead{Filtering irrelevant frequencies} Input time series may contain station-specific narrow-band noise that repeats over time. Similar time series segments generated by noise can be identified as near neighbors, or earthquake candidates in the similarity search. 

To suppress false positives generated from noise, we apply a bandpass filter to exclude frequency bands that show high average amplitudes and repeating patterns while containing low seismic activities. The bandpass filter is selected manually by examining short spectrogram samples, typically an hour long, of the input time series, based on seismological knowledge. Typical bandpass filter ranges span from 2 to 20Hz. Prior work~\cite{FAST, FASTlarge, FASTFingerPrint, networkpaper} proposes the idea of filtering irrelevant frequencies, but only on input time series. We extend the filter to the fingerprinting algorithm and cutoff spectrograms at the corner of the bandpass filter, which empirically improves detection performance. We perform a quantitative evaluation of the impact of bandpass filters on both the runtime and result quality (Section~\ref{eval:domain}).

\minihead{Removing correlated noise} Repeating non-seismic signals can also occur in frequency bands containing rich earthquake signals. Figure~\ref{fig:ltz_noise} shows an example of strong repeating background signals from a New Zealand seismic station. A large cluster of repeating signals with high pairwise similarity could produce nearest neighbor matches that dominate the similarity search, leading to a 10$\times$ increase in runtime and an over 100$\times$ increase in output size compared to results from similar stations. This poses both problems for computational scalability and for seismological interpretability.

We develop an occurrence filter for the similarity search by exploiting the rarity of the earthquake signals. Specifically, if a specific fingerprint is generating too many nearest neighbor matches in a short duration of time, we can be fairly confident that it is not an earthquake signal. This observation holds in general except for special scenarios such as volcanic earthquakes~\cite{volcanic}.

During the similarity search, we dynamically generate a list of fingerprints to exclude from future search. If the number of near neighbor candidates a fingerprint generates is larger than a predefined percentage of the total fingerprints, we exclude this fingerprint as well as its neighbors from future similarity search. To capture repeating noise over a short duration of time, the filter can be applied on top of the partitioned search. In this case, the filtering threshold is defined as the percentage of fingerprints in the current partition, rather than in the whole dataset. On the example dataset above, this approach filtered out around 30\% of the total fingerprints with no false positives. We evaluate the effect of the occurrence filter on different datasets under different filtering thresholds in Section~\ref{eval:domain}.

\section{Spatiotemporal Alignment}
\label{sec:network}
The LSH-based similar search outputs pairs of similar fingerprints (or waveforms) from the input, without knowing whether or not the pairs correspond to actual earthquake events. In this section, we show that by incorporating domain knowledge, we are able to significantly reduce the size of the output and prioritize seismic findings in the similarity search results. We briefly summarize the aggregation and filtering techniques on the level of seismic channels, seismic stations and seismic networks introduced in a recent paper in seismology~\cite{networkpaper} (Section~\ref{sec:networkoverview}). We then describe the implementation challenges and our out-of-core adaptations enabling the algorithm to scale to large output volumes (Section~\ref{sec:networkimp}).
\subsection{Alignment Overview}
\label{sec:networkoverview}
\begin{figure}[t!]
    \centering
    \includegraphics[width=\linewidth]{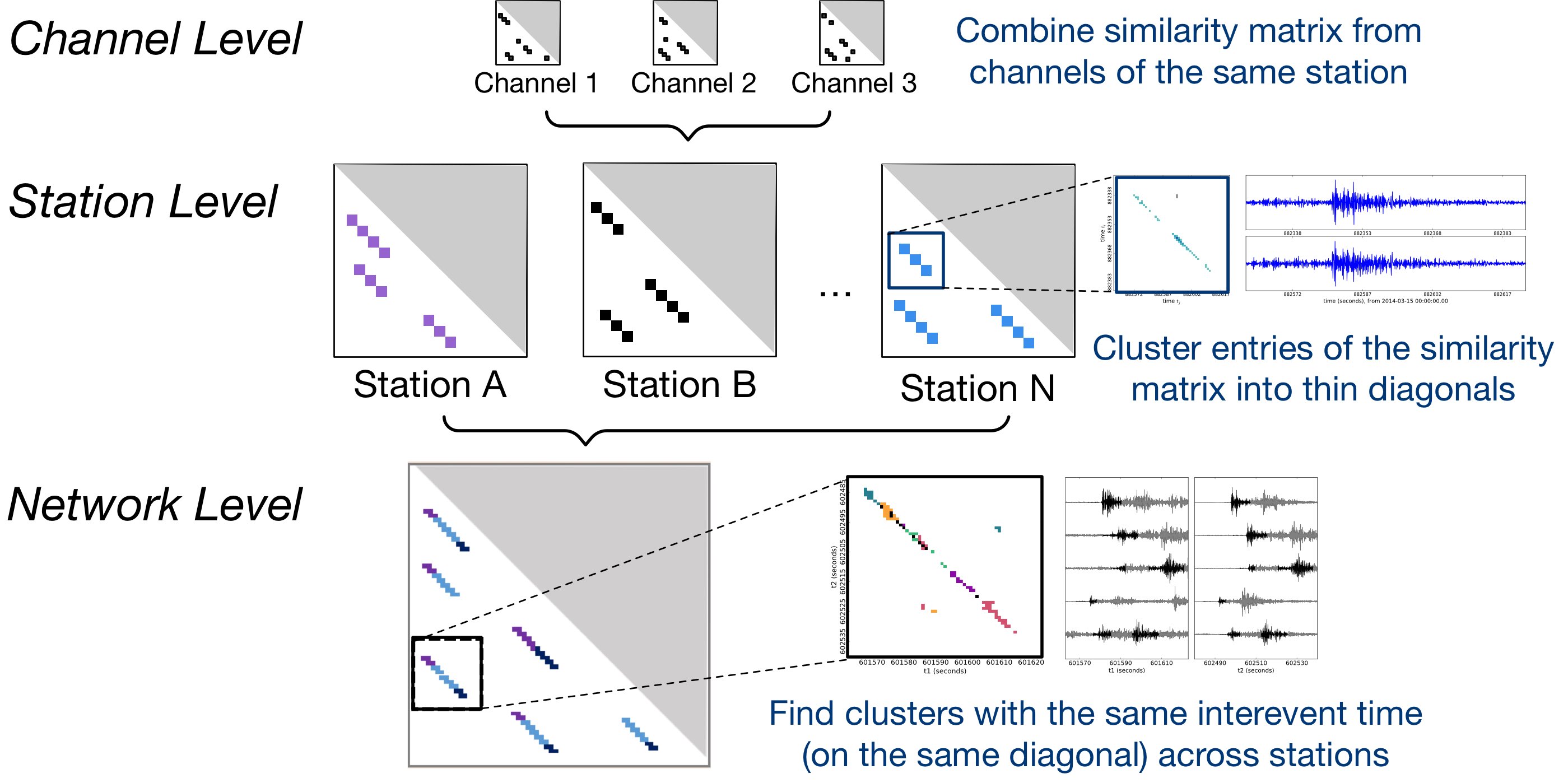}
    \vspace{-1em}
    \caption{The alignment procedure combines similarity search outputs from all channels in the same station (Channel Level), groups similar fingerprint matches generated from the same pair of reoccurring earthquakes (Station Level), and checks across seismic stations to reduce false positives in the final detection list (Network Level).}
    \label{fig:network_example}
\end{figure}

The similarity search computes a sparse similarity matrix $\mathcal{M}$, where the non-zero entry $\mathcal{M}[i, j]$ represents the similarity of fingerprints $i$ and $j$. In order to identify weak events in low signal-to-noise ratio settings, seismologists set lenient detection thresholds for the similarity search, resulting in large outputs in practice. For example, one year of input time series data can easily generate 100G of output, or more than 5 billion pairs of similar fingerprints. Since it is infeasible for seismologists to inspect all results manually, we need to automatically filter and align the similar fingerprint pairs into a list of potential earthquakes with high confidence. Based on algorithms proposed in a recent work in seismology~\cite{networkpaper}, we seek to reduce similarity search results at the level of seismic channels, stations and also across a seismic network. Figure~\ref{fig:network_example} gives an overview of the spatiotemporal alignment procedure. 

\minihead{Channel Level} Seismic channels at the same station experience ground movements at the same time. Therefore, we can directly merge detection results from each channel of the station by summing the corresponding similarity matrix. Given that earthquake-triggered fingerprint matches tend to register at multiple channels whereas matches induced by local noise might only appear on one channel, we can prune detections by imposing a slightly higher similarity threshold on the combined similarity matrix. This is to make sure that we include either matches with high similarity, or weaker matches registered at more than one channel.

\minihead{Station Level} Given a combined similarity matrix for each seismic station, domain scientists have found that earthquake events can be characterized by thin diagonal shaped clusters in the matrix, which corresponds to a group of similar fingerprint pairs separated by a constant offset~\cite{networkpaper}. The constant offset represents the time difference, or the inter-event time, between a pair of reoccurring earthquake events. One pair of reoccurring earthquake events can generate multiple fingerprint matches in the similarity matrix, since event waveforms are longer than a fingerprint time window. We exclude ``self-matches" generated from adjacent/overlapping fingerprints that are not attributable to reoccurring earthquakes. After grouping similar fingerprint pairs into clusters of thin diagonals, we reduce each cluster to a few summary statistics, such as the bounding box of the diagonal, the total number of similar pairs in the bounding box, and the sum of their similarity. Compared to storing every similar fingerprint pair, the clusters and summary statistics significantly reduce the size of the output.

\begin{figure}[t!]
    \centering
    \includegraphics[width=\linewidth]{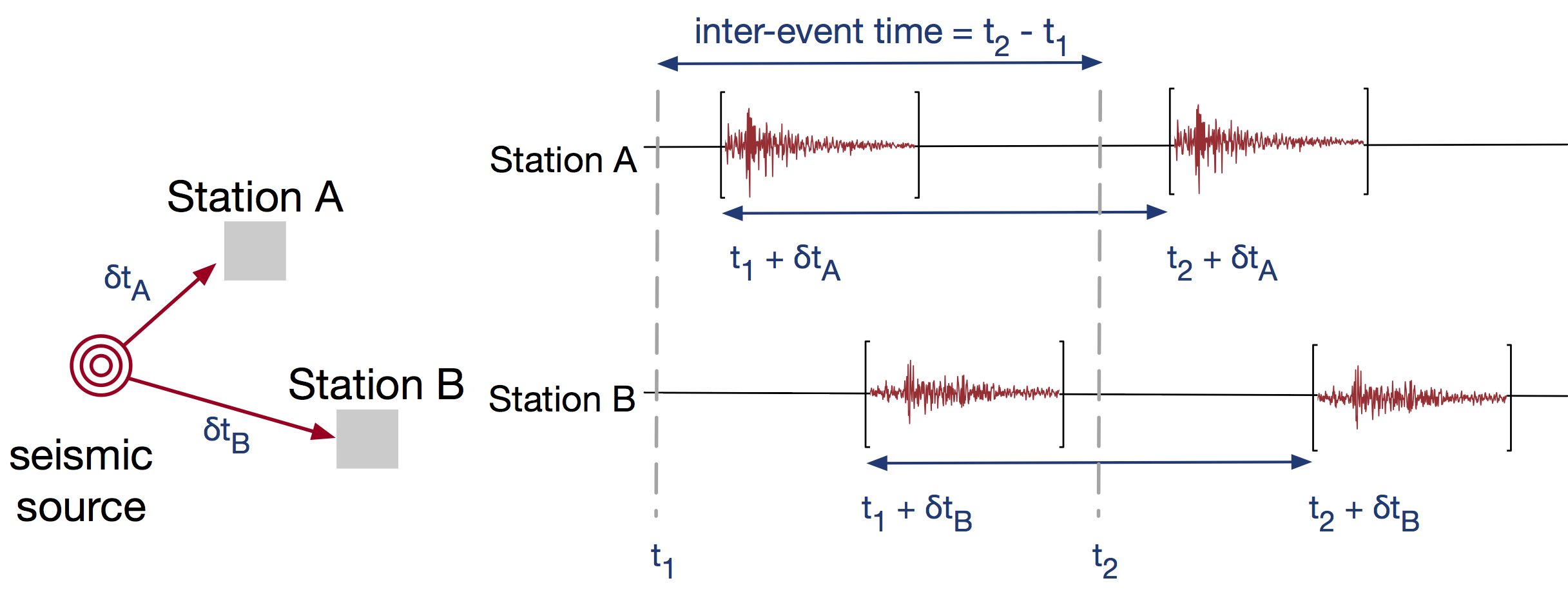}
    \vspace{-1.5em}
    \caption{Earthquakes from the same seismic sources has a fixed travel time to each seismic station (e.g. $\delta t_A$, $\delta t_B$ in the figure). The inter-event time between two occurrences of the same earthquake is invariant across seismic stations.}
    \label{fig:interevent}
\end{figure}

\minihead{Network Level} Earthquake signals also show strong temporal correlation across the seismic network, which we exploit to further suppress non-earthquake matches. Since an earthquake's travel time is only a function of its distance from the source but not of the magnitude, reoccurring earthquakes generated from the same source take a fixed travel time from the source to the seismic stations on each occurrence. Assume that an earthquake originated from source $X$ takes $\delta t_A$ and $\delta t_B$ to travel to seismic stations $A$ and $B$ and that the source generates two earthquakes at time $t_1$ and $t_2$ (Figure~\ref{fig:interevent}). Station $A$ experiences the arrivals of the two earthquakes at time $t_1 + \delta t_A$ and $t_2 + \delta t_A$, while station $B$ experiences the arrivals at $t_1 + \delta t_B$ and $t_2 + \delta t_B$. The inter-event time $\Delta t$ of these two earthquake events is independent of the location of the stations:
\[\Delta t = (t_2 + \delta t_A) - (t_1 + \delta t_A) = (t_2 + \delta t_B) - (t_1 + \delta t_B) = t_2 - t_1.\]
This means that in practice, diagonals with the same offset $\Delta t$ and close starting times at multiple stations can be attributed to the same earthquake event. We require a pair of earthquake events to be observed at more than a user-specified number of stations in order to be considered as a detection.

On a run with 7 to 10 years of time series data from 11 seismic stations (27 channels), the postprocessing procedure effectively reduced the output from more than 2 Terabytes of similar fingerprint pairs to around 30K timestamps of potential earthquakes.

\subsection{Implementation and Optimization}
\label{sec:networkimp}
The volume of similarity search output poses serious challenges for the alignment procedure, as we often need to process results larger than the main memory of a single node. In this subsection, we describe our implementation and the new out-of-core adaptations of the algorithm that enable the scaling to large output volumes.

\minihead{Similarity search output format} The similarity search produces outputs that are in the form of triplets. A triplet $(dt, idx1, sim)$ is a non-zero entry in the similarity matrix, which represents that fingerprint $idx1$ and $(idx1 + dt)$ are hashed into the same bucket $sim$ times (out of $t$ independent trials). We use $sim$ as an approximation of the similarity between the two fingerprints.

\minihead{Channel} First, given outputs of similar fingerprint pairs (or the non-zero entries of the similarity matrix) from different channels at the same station, we want to compute the combined similarity matrix with only entries above a predefined threshold.

Na\"ively, we could update a shared hashmap of the non-zero entries of the similarity matrix for each channel in the station. However, since the hashmap might not fit in the main memory on a single machine, we utilize the following sort-merge-reduce procedure instead:
\begin{enumerate}[noitemsep,topsep=.5em]
    \item In the sorting phase, we perform an external merge sort on the outputs from each channel, with $dt$ as the primary sort key and $idx1$ as the secondary sort key. That is, we sort the similar fingerprint pairs first by the diagonal that they belong to in the similarity matrix, and within the diagonals, by the start time of the pairs.
    \item In the merging phase, we perform a similar external merge sort on the already sorted outputs from each channel. This is to make sure that all matches generated by the same pair of fingerprint $idx1$ and $idx1 + dt$ at different channels can be concentrated in consecutive rows of the merged file.
    \item In the reduce phase, we traverse through the merged file and combine the similarity score of consecutive rows of the file that share the same $dt$ and $idx1$. We discard results that have combined similarity smaller than the threshold.
\end{enumerate}

\minihead{Station} Given a combined similarity matrix for each seismic station, represented in the form of its non-zero entries sorted by their corresponding diagonals and starting time, we want to cluster fingerprint matches generated by potential earthquake events, or cluster non-zero entries along the narrow diagonals in the matrix.

We look for sequences of detections (non-zero entries) along each diagonal $dt$, where the largest gap between consecutive detections is smaller than a predefined gap parameter. Empirically, permitting a gap help ensure an earthquake's P and S wave arrivals are assigned to the same cluster. Identification of the initial clusters along each diagonal $dt$ requires a linear pass through the similarity matrix. We then interactively merge clusters in adjacent diagonals $dt-1$ and $dt+1$, with the restriction that the final cluster has a relatively narrow width. We store a few summary statistics for each cluster (e.g. the cluster's bounding box, the total number of entries) as well as prune small clusters and isolated fingerprint matches, which significantly reduces the output size.

The station level clustering dominates the runtime in the spatiotemporal alignment. In order to speed up the clustering, we partition the similarity matrix according to the diagonals, or ranges of $dt$s of the matched fingerprints, and perform clustering in parallel on each partition. A na\"ive equal-sized partition of the similarity matrix could lead to missed detections if a cluster split into two partitions gets pruned in both due to the decrease in size. Instead, we look for proper points of partition in the similarity matrix where there is a small gap between neighboring occupied diagonals. Again, we take advantage of the ordered nature of similarity matrix entries. We uniformly sample entries in the similarity matrix, and for every pair of neighboring sampled entries, we only check the entries in between for partition points if the two sampled entries lie on diagonals far apart enough to be in two partitions. Empirically, a sampling rate of around 1\% works well for our datasets in that most sampled entries are skipped because they are too close to be partitioned.

\minihead{Network} Given groups of potential events at each station, we perform a similar summarization across the network in order to identify subsets of the events that can be attributed to the same seismic source. In principle, we could also partition and parallelize the network detection. In practice, however, we found that the summarized event information at each station is already small enough that it suffices to compute in serial.

\section{Evaluation}
\label{sec:eval}

In this section, we perform both quantitative evaluation on performances of the detection pipeline, as well as qualitative analysis of the detection results. Our goal is to demonstrate that:
\begin{enumerate}[topsep=.5em]
  \setlength\itemsep{0.2em}
    \item Each of our optimizations contributes meaningfully to the performance improvement; together, our optimizations enable an over 100$\times$ speed up in the end-to-end pipeline. 
    \item Incorporating domain knowledge in the pipeline improves both the performance and the quality of the detection.
    \item The improved scalability enables scientific discoveries on two public datasets: we discovered 597 new earthquakes from a decade of seismic data near the Diablo Canyon nuclear power plant in California, as well as 6123 new earthquakes from a year of seismic data from New Zealand.
\end{enumerate}

\minihead{Dataset} We evaluate on two public datasets used in seismological analyses with our domain collaborators. The first dataset includes 1 year of 100Hz time series data (3.15 billion points per station) from 5 seismic stations (LTZ, MQZ, KHZ, THZ, OXZ) in New Zealand. We use the vertical channel (usually the least noisy) from each station~\cite{GeoNet}. The second dataset of interest includes 7 to 10 years of 100Hz time series data from 11 seismic stations and 27 total channels near the Diablo Canyon power plant in California~\cite{NCEDC}.

\minihead{Experimental Setup} We report results from evaluating the pipeline on a server with 512GB of RAM and two 28-thread Intel Xeon E5-2690 v4 2.6GHz CPUs. Our test server has L1, L2, L3 cache sizes of 32K, 256K and 35840K. We report the runtime averages from multiple trials.

\subsection{End-to-end Evaluation}
\label{sec:e2e}
In this subsection, we report the runtime breakdown of the baseline implementation of the pipeline, as well as the effects of applying different optimizations.

\begin{figure*}
    \centering
    \includegraphics[width=0.95\linewidth]{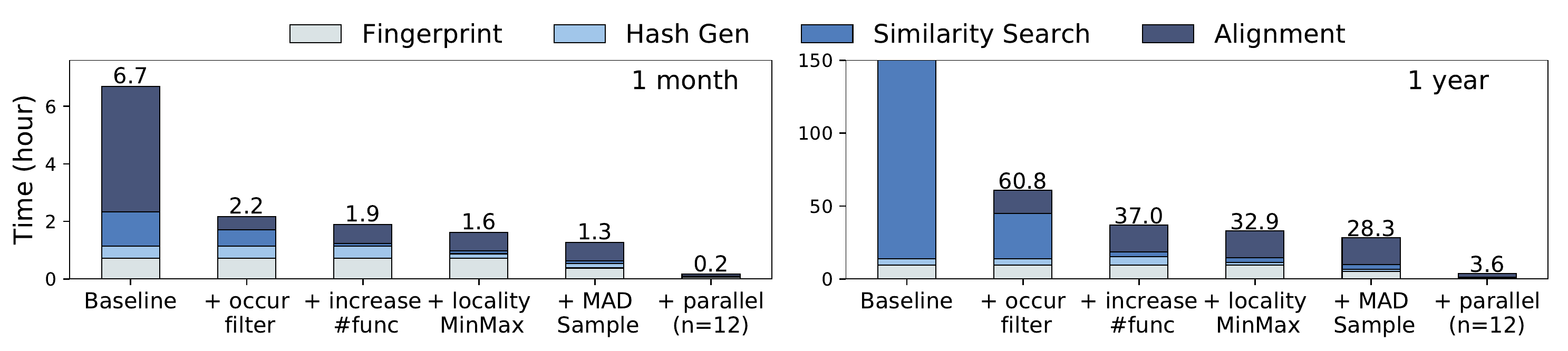}
    \vspace{-1em}
    \caption{Factor analysis of processing 1 month (left) and 1 year (right) of 100Hz data from LTZ station in the New Zealand dataset. We show that each of our optimization contributes to the performance improvements, and enabled an over 100$\times$ speed up end-to-end. }
    \label{fig:e2etime}
\end{figure*}

To evaluate how our optimizations scale with data size, we evaluate the end-to-end pipeline on 1 month and 1 year of time series data from station LTZ in the New Zealand dataset. We applied a bandpass filter of 3-20Hz on the original time series to exclude noisy low-frequency bands. For fingerprinting, we used a sliding window with length of 30 seconds and slide of 2 seconds, which results in 1.28M binary fingerprints for 1 month of time series data (15.7M for one year), each of dimension 8192; for similarity search, we use $6$ hash functions, and require a detection threshold of $5$ matches out of $100$ hash tables. We further investigate the effect of varying these parameters in the microbenchmarks in Section~\ref{eval:params}.

Figure~\ref{fig:e2etime} shows the cumulative runtime after applying each optimization. Overall, our optimizations scale well with the size of the dataset, and enable an over 100$\times$ improvement in end-to-end processing time. We analyze each of these components in turn:

First, we apply a 1\% occurrence filter (+ occur filter, Section~\ref{sec:noise}) during similarity search to exclude frequent fingerprint matches generated by repeating background noise. This enables a 2-5$\times$ improvement in similarity search runtime while reducing the output size by 10-50$\times$, reflected in the decrease in postprocessing time.

Second, we further reduce the search time by increasing the number of hash functions to 8 and lowering the detection threshold to 2 (+ increase \#funcs, Section~\ref{sec:searchparam}). While this increases the hash signature generation and output size, it enables around 10$\times$ improvement in search time for both datasets. 

Third, we reduce the hash signature generation time by improving the cache locality and reducing the computation with Min-Max hash instead of \mh (+ locality MinMax, Section~\ref{sec:hashgen}), which leads to a 3$\times$ speedup for both datasets.

Fourth, we speed up fingerprinting by 2$\times$ by estimating MAD statistics with a 10\% sample (+ MAD sample, Section~\ref{sec:mad}).

Finally, we enable parallelism and run the pipeline with 12 threads (Section~\ref{sec:mad},~\ref{sec:searchpart},~\ref{sec:networkimp}). As a result, we see an almost linear decrease in runtime in each part of the pipeline. Notably, due to the overall lack of data dependencies in this scientific pipeline, simple parallelization can already enable significant speedups. 

The improved scalability enables us to scale analytics from 3 months to over 10 years of data. We discuss qualitative detection results from both datasets in Section~\ref{sec:eq}.

\subsection{Effect of domain-specific optimizations}
\label{eval:domain}
In this section, we investigate the effect of applying domain-specific optimizations to the pipeline. We demonstrate that incorporating domain knowledge could improve both performance and result quality of the detection pipeline.

\minihead{Occurrence filter} We evaluate the effect of applying the occurrence filter during similarity search on the five stations from the New Zealand dataset. For this experiment, we use a partition size of 1 month as the duration for the occurrence threshold; a $>$1\% threshold indicates that a fingerprint matches over 1\% (10K) other fingerprints in the same month. We report the total percentage of filtered fingerprints under varying thresholds in Table~\ref{tab:missedevents}. We also evaluate the accuracy of the occurrence filter by comparing the timestamps of filtered fingerprints with the catalog of the arrival times of known earthquakes at each station. We report the false positive rate, or the number of filtered earthquakes over the total number of cataloged events, of the filter under varying thresholds.

The results show that as the occurrence filter becomes stronger, the percentage of filtered fingerprints and the false positive rate both increase. For seismic stations suffering from correlated noise, the occurrence filter can effectively eliminate a significant amount of fingerprints from the similarity search. For station LTZ, a $>$1\% threshold filters out up to 30\% of the total fingerprints without any false positives, which results in a 4$\times$ improvement in runtime. For other stations, the occurrence filter has little influence on the results. This is expected since these stations do not have repeating noise signals present at station LTZ (Figure~\ref{fig:ltz_noise}). In practice, correlated noise is rather prevalent in seismic data. In the Diablo Canyon dataset for example, we applied the occurrence filter on three out of the eleven seismic stations in order for the similarity search to finish in a tractable time.  


\begin{table*}
\centering
\small
\ra{1.1}
\begin{tabular}{@{}rrrrcrrrcrrrcrrrcrrr@{}}
\hlineB{1.5}
& \multicolumn{3}{c}{\textbf{LTZ} (1548 events)} & \phantom{a}& \multicolumn{3}{c}{\textbf{MQZ} (1544 events)} &
\phantom{a} & \multicolumn{3}{c}{\textbf{KHZ} (1542 events)} &
\phantom{a} & \multicolumn{3}{c}{\textbf{THZ} (1352 events)}&
\phantom{a} & \multicolumn{3}{c}{\textbf{OXZ} (1248 events)}\\
\cmidrule{2-4} \cmidrule{6-8} \cmidrule{10-12}  \cmidrule{14-16} \cmidrule{18-20}
\textbf{Thresh} & FP & Filtered & Time && FP & Filtered& Time&&  FP& Filtered & Time && FP& Filtered & Time &&  FP& Filtered& Time\\ 
\hline
$>$5.0\%   & 0 & 0.09 & 149.3 && 0 & 0 & 2.8 && 0& 0 & 2.2 &&  0& 0 & 2.4 && 0& 0 & 2.6\\
$>$1.0\%   & 0 & 30.1 & 31.0 && 0 & 0 & 2.7 && 0& 0 & 2.3 &&  0& 0 &  2.3 && 0& 0 & 2.6\\
$>$0.5\% & 0 & 31.2 & 32.1 && 0 & 0.09 & 2.8 && 0& 0 & 2.4 &&  0& 0 & 2.4 && 0.08 & 0.08 & 2.7\\
$>$0.1\% & 0 & 32.1 & 28.6 && 0.07 & 0.3 & 2.7 && 0 & 0.03 & 2.4 && 0& 0.02 & 2.3 && 0.08& 0.17 & 2.6\\
\hlineB{1.5}
\end{tabular}
\vspace{0.5em}
\caption{The table shows that the percentage of fingerprints filtered (Filtered) and the false positive rate (FP) both increase as the occurrence filter becomes stronger (from filtering matches above 5.0\% to above 0.1\%). The runtime (in hours) measures similarity search time. }
\label{tab:missedevents}
\end{table*}

\begin{figure}
    \centering
    \includegraphics[width=\linewidth]{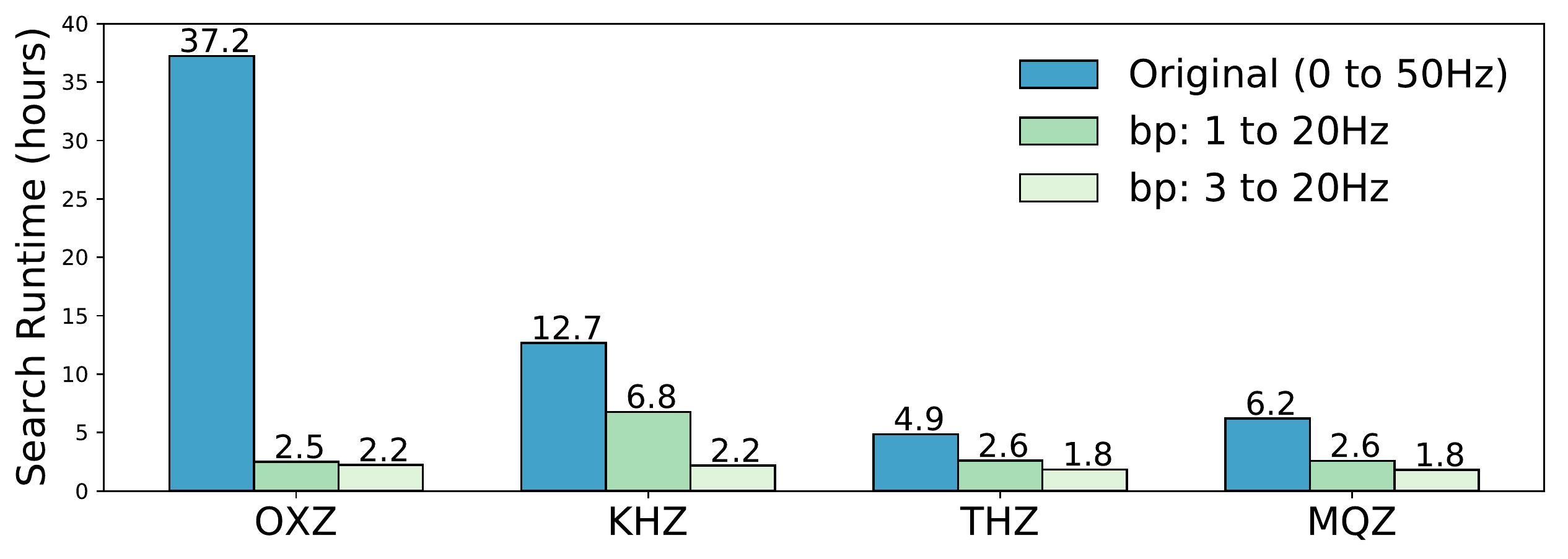}
    \vspace{-1.1em}
    \caption{LSH runtime under different band pass filters. Matches of noise in the non-seismic frequency bands can lead to a 16$\times$ increase in runtime and over 200 $\times$ increase in output size for unfiltered time series.}
    \label{fig:bpfilter}
\end{figure}

\minihead{Bandpass filter} We compare similarity search on the same dataset (Nyquist frequency 50Hz) before and after applying bandpass filters. The first bandpass filter (bp: 1-20Hz) is selected as most seismic signals are under 20Hz; the second (bp: 3-20Hz) is selected after manually looking at samples spectrograms of the dataset and excluding noisy low frequencies. Figure~\ref{fig:bpfilter} reports the similarity search runtime for fingerprints generated with different bandpass filters. Overall, similarity search suffers from additional matches generated from the noisy frequency bands outside the interests of seismology. For example, at station OXZ, removing the bandpass filter leads to a 16$\times$ slow down in runtime and a 209$\times$ increase in output size.

We compare detection recall on 8811 catalog earthquake events for different bandpass filters. The recall for the unfiltered data (0-50Hz), the 1-20Hz and 3-20Hz bandpass filters are 20.3\%, 23.7\%, 45.2\%, respectively. The overall low recall is expected, as we only used 4 (out of over 50) stations in the seismic network that contributes to the generation of catalog events. Empirically, a narrow, domain-informed bandpass filter focuses the comparison of fingerprint similarity on frequencies that are characteristics of seismic events, leading to improved similarity between earthquake events and therefore increased recall. We provide guidelines for setting the bandpass filter in Appendix~\ref{appendix:bp}. 

\subsection{Effect of pipeline parameters}
\label{eval:params}

In this section, we evaluate the effect of the space/quality and time trade-offs for core pipeline parameters.

\minihead{MAD sampling rate} We evaluate the speed and quality trade-off for calculating the median and MAD of the wavelet coefficients for fingerprints via sampling. We measure the runtime and accuracy on the 1 month dataset in Section~\ref{sec:e2e} (1.3M fingerprints) under varying sampling rates. Overall, runtime and accuracy both decrease with sampling rate as expected. For example, a 10\% and 1\% sampling rate produce fingerprints with 99.7\% and 98.7\% accuracy, while enabling a near linear speedup of 10.5$\times$ and 99.8$\times$, respectively. Below 1\%, runtime improvements suffer from a diminishing return, as the IO begins to dominate the MAD calculation in runtime--on this dataset, a 0.1\% sampling rate only speeds up the MAD calculation by 350$\times$. We include additional results of this trade-off in the appendix.

\begin{figure}
    \centering
    \includegraphics[width=\linewidth]{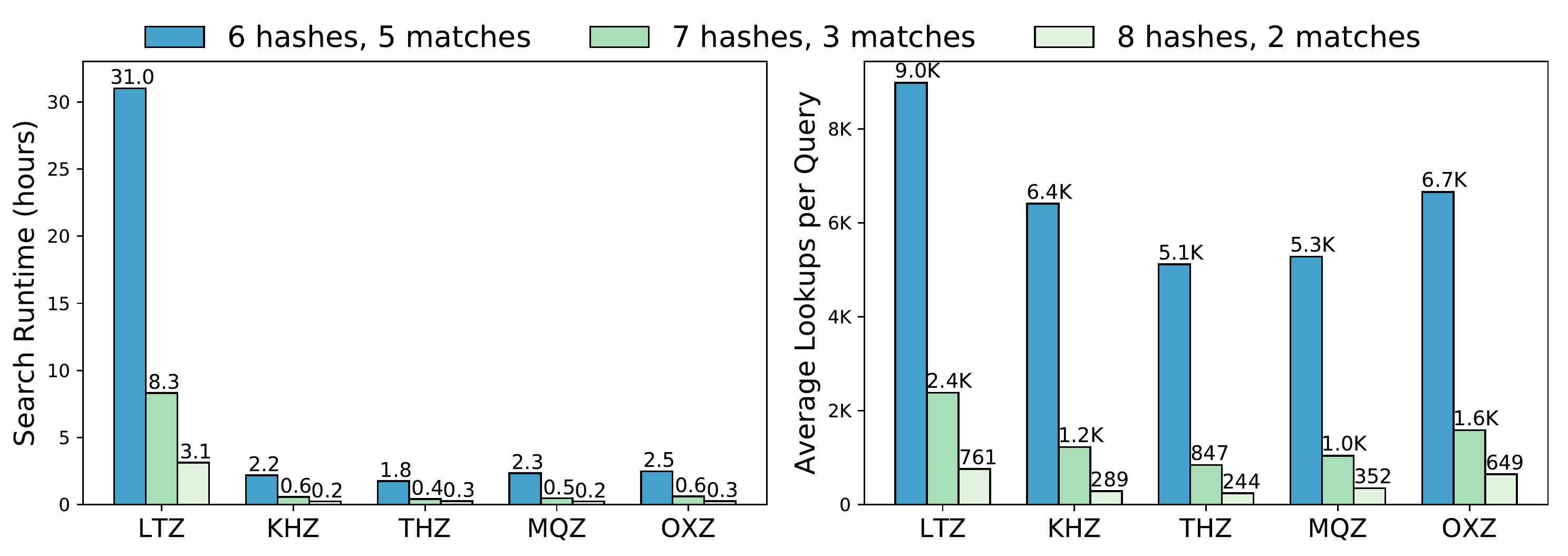}
    \vspace{-1.5em}
    \caption{Effect of LSH parameters on similarity search runtime and average query lookups. Increasing the number of hash functions significantly decreases average number of lookups per query, which results in an up to 10$\times$ improvement in runtime. }
    \label{fig:lshparam}
\end{figure}

\minihead{LSH parameters} We report runtime of the similarity search under different LSH parameters in Figure~\ref{fig:lshparam}. As indicated in Figure~\ref{fig:prob}, the three sets of parameters that we evaluate yield near identical probability of detection given Jaccard similarity of two fingerprints. However, by increasing the number of hash functions and thereby increasing the selectivity of hash signatures, we decrease the average number of lookups per query by over 10x. This results in around 10x improvement in similarity search time. 

\begin{figure}
    \centering
    \includegraphics[width=0.95\linewidth]{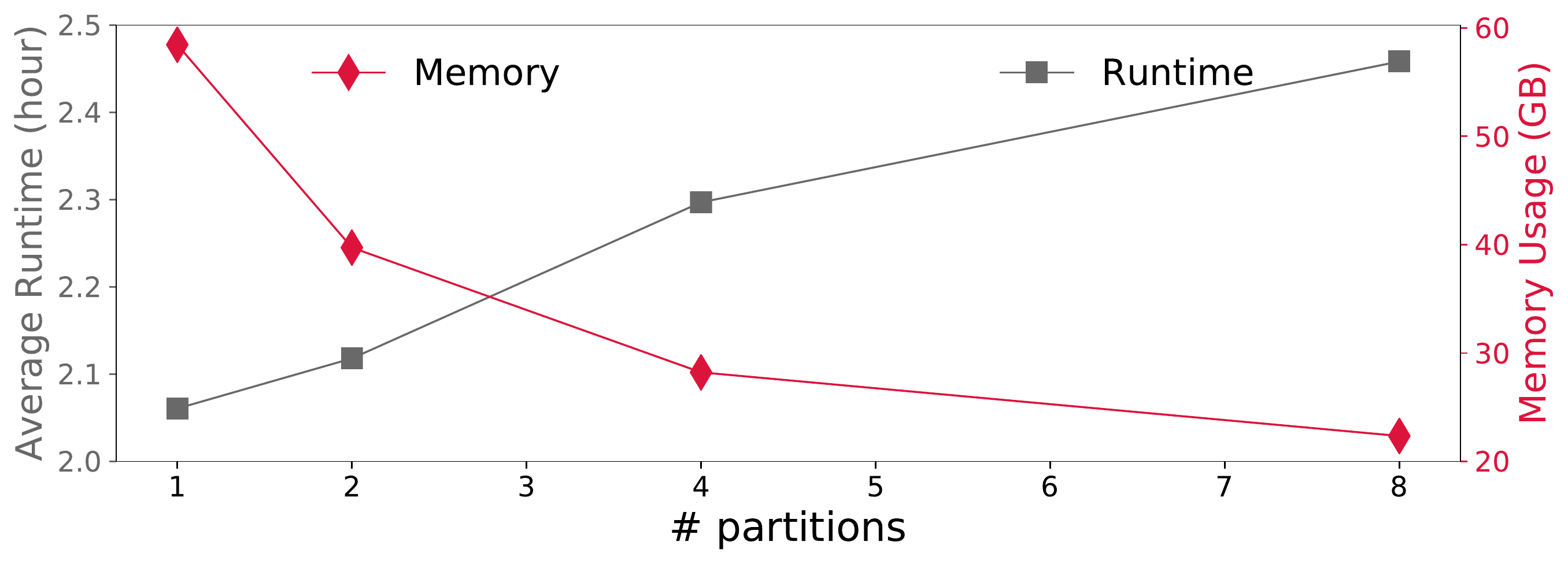}
    \vspace{-0.5em}
    \caption{Runtime and memory usage for similarity search under a varying number of partitions. By increasing the number of search partitions, we are able to decrease the memory usage by over 60\% while incurring less than 20\% runtime overhead.}
    \label{fig:partition}
\end{figure}

\minihead{Number of partitions} We report the runtime and memory usage of the similarity search with varying number of partitions in Figure~\ref{fig:partition}. As the number of partitions increases, the runtime increases slightly due to the overhead of initialization and deletion of hash tables. In contrast, memory usage decreases as we only need to keep a subset of the hash signatures in the hash table at any time. Overall, by increasing the number of partitions from 1 to 8, we are able to decrease the memory usage by over 60\% while incurring less than 20\% runtime overhead. This allows us to run LSH on larger datasets with the same amount of memory. 

\begin{figure}
    \centering
    \includegraphics[width=0.95\linewidth]{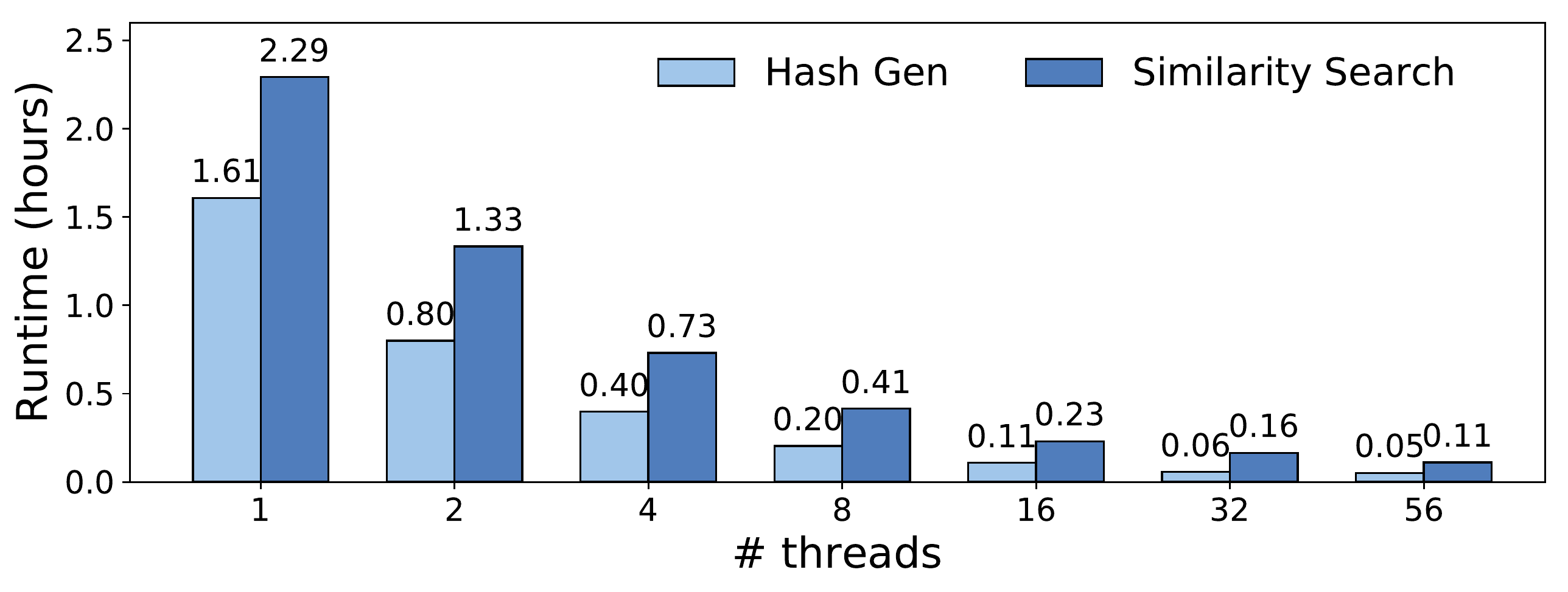}
    \vspace{-0.5em}
    \caption{Hash generation scales near linearly up to 32 threads. }
    \label{fig:parallel}
\end{figure}

\minihead{Parallelism} Finally, to quantify the speedups from parallelism, we report the runtime of LSH hash signature generation and similarity search using a varying number of threads. For hash signature generation, we report time taken to generate hash mappings as well as the time taken to compute Min-Max hash for each fingerprint. For similarity search, we fix the input hash signatures and vary the number of threads assigned during the search. We show the runtime averaged from four seismic stations in Figure~\ref{fig:parallel}. Overall, hash signature generation scales almost perfectly (linearly) up to 32 threads, while similarity search scales slightly worse; both experience significant performance degradation running with all available threads.

\begin{figure*}
    \centering
    \includegraphics[width=\linewidth]{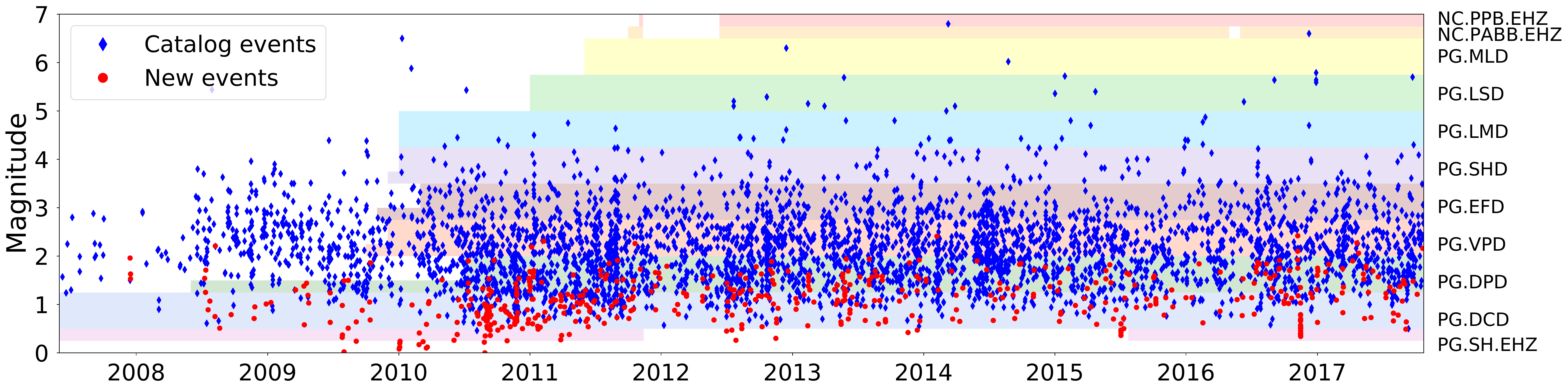}
    \vspace{-1.8em}
    \caption{The left axis shows origin times and magnitude of detected earthquakes, with the catalog events marked in blue and new events marked in red. The colored bands in the right axis represent the duration of data used for detection collected from 11 seismic stations and 27 total channels. Overall, we detected 3957 catalog earthquakes (diamond) as well as 597 new local earthquakes (circle) from this dataset.}
    \label{fig:neweq}
\end{figure*}

\subsection{Comparison with Alternatives}
\label{sec:falconn}
In this section, we evaluate against alternative similarity search algorithms and supervised methods. We include additional experiment details in Appendix~\ref{appendix:alternate}.

\minihead{Alternative Similarity Search Algorithms} We compare the single-core query performance of our MinHash LSH to 1) an alternative open source LSH library FALCONN~\cite{falconnlib} 2) four state-of-the-art set similarity join algorithms: PPJoin~\cite{ppjoin}, GroupJoin~\cite{groupjoin}, AllPairs~\cite{allpairs} and AdaptJoin~\cite{adaptjoin}. We use 74,795 fingerprints with dimension 2048 and 10\% non-zero entries, and a Jaccard similarity threshold of 0.5 for all libraries. Compared to exact algorithms like set similarity joins, approximate algorithms such as LSH incur a 6\% false negative rate. However, MinHash LSH enables a 24$\times$ to 65$\times$ speedup against FALCONN and 63$\times$ to 197$\times$ speedup against set similarity joins (Table~\ref{tab:setsim}). Characteristics of the input fingerprints contribute to the performance differences: the fixed number of non-zero entries in fingerprints makes pruning techniques in set similarity joins based on set length irrelevant; our results corroborate with previous findings that \mh outperforms SimHash on binary, sparse input~\cite{minhashsimhash}.

\begin{table}
\small \center
\begin{tabular}{r r r}
\hlineB{1.5}
\textbf{Algorithm} & \textbf{Average Query time} & \textbf{Speedup}\\
\hline
MinHash LSH & 36 $\mu$s& -- \\
FALCONN vanilla LSH & .87ms & 24$\times$ \\
FALCONN multi-probe LSH & 2.4ms & 65$\times$\\
AdaptJoin~\cite{adaptjoin} & 2.3ms& 63$\times$ \\
AllPairs~\cite{allpairs} & 7.1ms& 197$\times$ \\
GroupJoin~\cite{groupjoin} & 5.7ms& 159$\times$ \\
PPJoin~\cite{ppjoin} & 5.5ms& 151$\times$ \\
\hlineB{1.5}
\end{tabular}
\vspace{0.5em}
\caption{Single core per-datapoint query time for LSH and set similarity joins. MinHash LSH incurs a 6.6\% false negative rate while enabling up to 197$\times$ speedup. }
\label{tab:setsim}
\end{table}

\minihead{Supervised Methods} We report results evaluating two supervised models: WEASEL~\cite{weasel} and ConvNetQuake~\cite{convquake} on the Diablo Canyon dataset. Both models were trained on labeled catalog events  (3585 events from 2010 to 2017) and randomly sampled noise windows at station PG.LMD. We also augment the earthquake training examples by 1) adding earthquake examples from another station PG.DCD 2) perturbing existing events with white noise 3) shifting the location of the earthquake event in the window. 
Table~\ref{tab:model} reports test accuracy of the two models on a sample of 306 unseen catalog events and 449 new events detected by our pipeline (FAST events), as well as the false positive rate estimated from manual inspection of 100 random earthquake predictions. While supervised methods achieve high accuracy in classifying  unseen catalog and noise events, they exhibit a high false positive rate (90$\pm$5.88\%) and miss 30-32\% of new earthquake events detected by our pipeline. The experiment suggests that unsupervised methods like our pipeline are able to detect qualitatively different events from the existing catalog, and that supervised methods are complements, rather than replacements, of unsupervised methods for earthquake detection.
\begin{table}
\small\center
\begin{tabular}{ r r r}
\hlineB{1.5}
 & \textbf{WEASEL~\cite{weasel}} & \textbf{ConvNetQuake~\cite{convquake}} \\ 
\hline
Test Catalog Acc. (\%) & 90.8 & 90.6 \\
Test FAST Acc. (\%) & 68.0 & 70.5 \\
True Negative Rate (\%) & 98.6 & 92.2\\
False Positive Rate (\%) & 90.0$\pm$5.88 & 90.0$\pm$5.88 \\
\hlineB{1.5}
\end{tabular}
\vspace{0.5em}
\caption{Supervised methods trained on catalog events exhibit high false positive rate and a 20\% accuracy gap between predictions on catalog and FAST detected events.  }
\label{tab:model}
\end{table}

\subsection{Qualitative Results}
\label{sec:eq}
We first report our findings in running the pipeline over a decade (06/2007 to 10/2017) of continuous seismic data from 11 seismic stations (27 total channels) near the Diablo Canyon nuclear power plant in central California. The chosen area is of special interest as there are many active faults near the power plant. Detecting additional small earthquakes in this region will allow seismologists to determine the size and shape of nearby fault structures, which can potentially inform seismic hazard estimates.

We applied station-specific bandpass filters between 3 and 12 Hz to remove repeating background noise from the time series. In addition, we applied the occurrence filter on three out of the eleven seismic stations that experienced corrupted sensor measurements. The number of input binary fingerprints for each seismic channel ranges from 180 million to 337 million; the similarity search runtime ranges from 3 hours to 12 hours with 48 threads.

\begin{figure}
    \centering
    \includegraphics[width=0.95\linewidth]{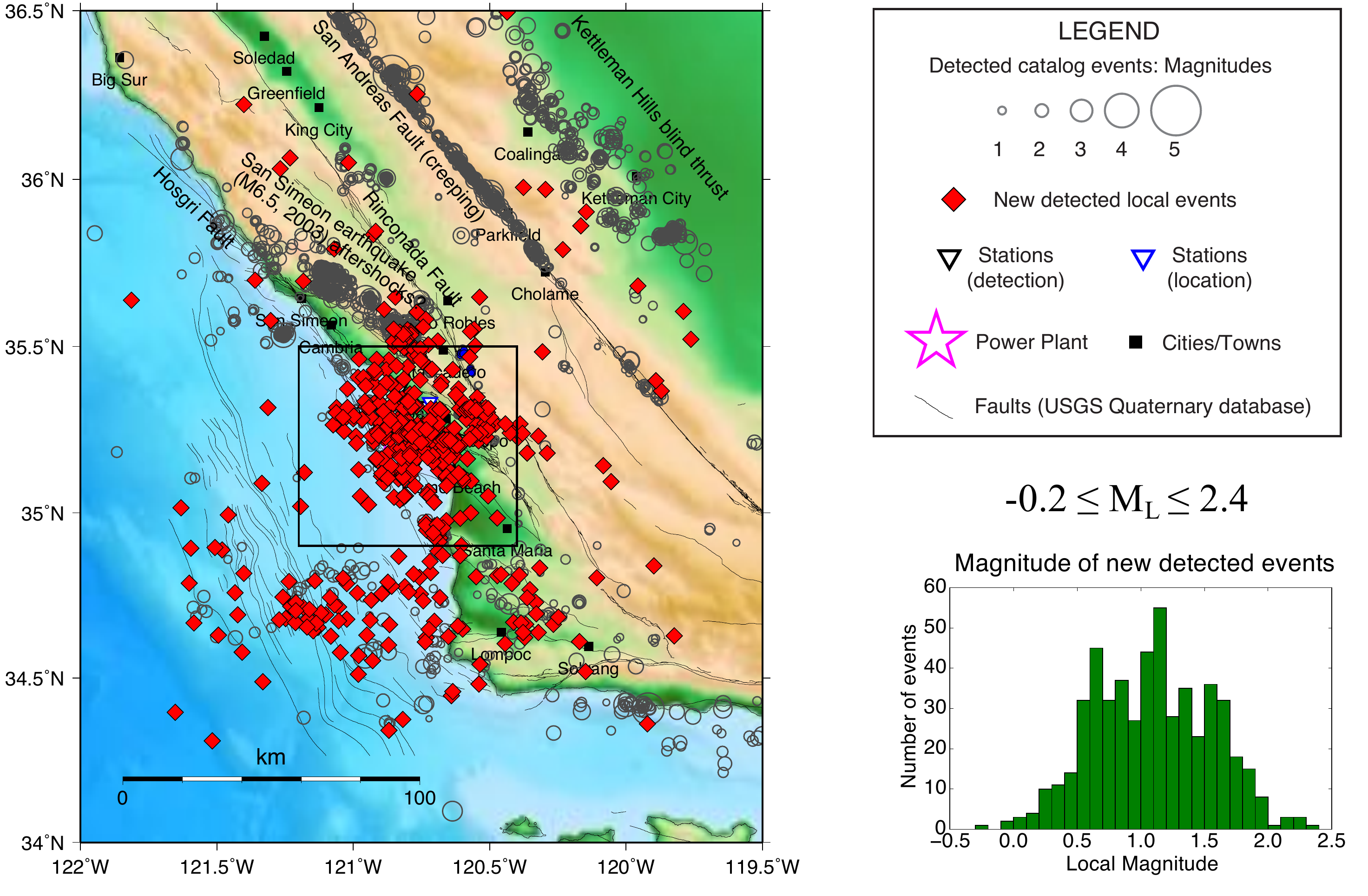}
    \vspace{-0.5em}
    \caption{Overview of the location of detected catalog events (gray open circles) and new events (red diamonds). The pipeline was able to detect earthquakes close to the seismic network (boxed) as well as all over California.}
    \label{fig:eqloc}
\end{figure}

Among the 5048 detections above our detection threshold, 397 detections (about 8\%) were false positives, confirmed via visual inspection: 30 were duplicate earthquakes with a lower similarity, 18 were catalog quarry blasts, 5 were deep teleseismic earthquakes (large earthquakes from $>$1000 km away). There were also 62 non-seismic signals detected across the seismic network; we suspect that some of these waveforms are sonic booms.

\begin{figure}
    \centering
    \includegraphics[width=0.95\linewidth]{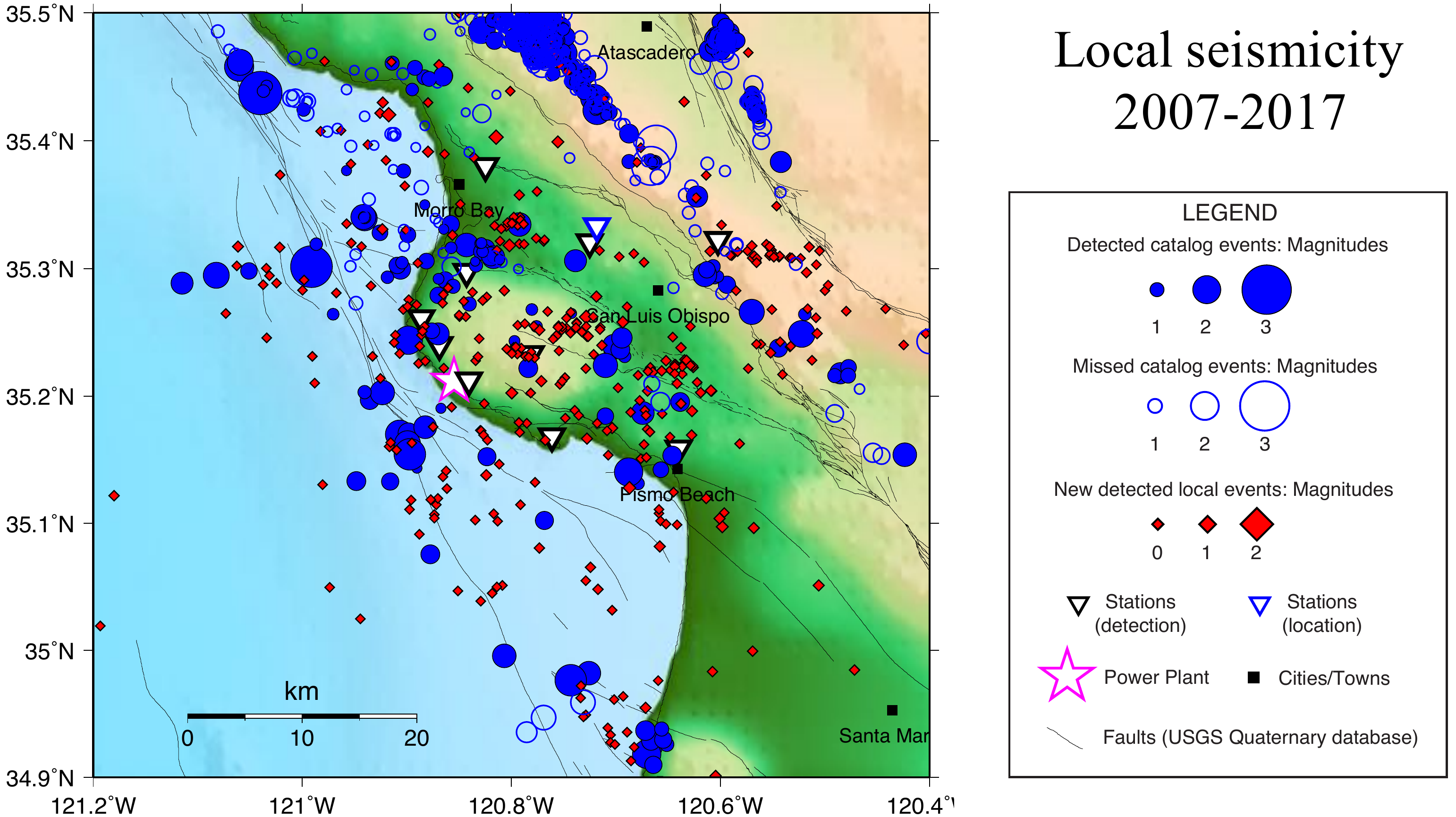}
    \vspace{-0.5em}
    \caption{Zoom in view of locations of new detected earthquakes (red diamonds) and cataloged events (blue circles) near the seismic network (box in Figure~\ref{fig:eqloc}). The new local earthquakes contribute detailed information about the structure of faults.}
    \label{fig:zoomineqloc}
\end{figure}

Overall, we were able to detect and locate 3957 catalog earthquakes, as well as 597 new local earthquakes. Figure~\ref{fig:neweq} shows an overview of the origin time of detected earthquakes, which is spread over the entire ten-year span. The detected events include both low-magnitude events near the seismic stations, as well as larger events that are farther away. Figure~\ref{fig:eqloc} visualizes the locations of both catalog events and newly detected earthquakes, and Figure~\ref{fig:zoomineqloc} zooms in on earthquakes in the vicinity of the power plant. Despite the low rate of local earthquake activity (535 total catalog events from 2007 to 2017 within the area shown in Figure~\ref{fig:zoomineqloc}), we were able to detect 355 new events that are between $-0.2$ and 2.4 in magnitude and located within the seismic network, where many active faults exist. We missed 261 catalog events, almost all of which originated from outside the network of our interest. Running the detection pipeline at scale enables scientists to discover earthquakes from unknown sources. These new detected events will be used to determine the details of active fault structures near the power plant.

We are also actively working with our domain collaborators on additional analysis of the New Zealand dataset. The pipeline detected 11419 events, including 4916 catalog events, 355 teleseismic events, 6123 new local earthquakes and 25 false positives (noise waveforms) verified by the seismologists. We are preparing these results for publication in seismological venues, and expect to further improve the detection results by scaling up the analysis to more seismic stations over a longer duration of time.

\section{Conclusion}

In this work, we reported on a novel application of LSH to large-scale seismological data, as well as the challenges and optimizations required to scale the system to over a decade of continuous sensor data. This experience in scaling LSH for large-scale earthquake detection illustrates both the potential and the challenge of applying core data analytics primitives to data-driven domain science on large datasets.
On the one hand, LSH and, more generally, time series similarity search, is well-studied, with scores of algorithms for efficient implementation: by applying canonical MinHash-based LSH, our seismologist collaborators were able to meaningfully analyze more data than would have been feasible via manual inspection.
On the other hand, the straightforward implementation of LSH in the original FAST detection pipeline failed to scale beyond a few months of data.
The particulars of seismological data---such as frequency imbalance in the time series and repeated background noise---placed severe strain on an unmodified LSH implementation and on researchers attempting to understand the output.
As a result, the seismological discoveries we have described in this paper would not have been possible without domain-specific optimizations to the detection pipeline.
We believe that these results have important implications for researchers studying LSH (e.g., regarding the importance of skew resistance) and will continue to bear fruit as we scale the system to even more data and larger networks.

\section*{Acknowledgements}
We thank the many members of the Stanford InfoLab for their valuable
feedback on this work. This research was supported in part by affiliate members and other supporters of the Stanford DAWN project---Facebook, Google, Intel, Microsoft, NEC, SAP, Teradata, and VMware---as well as Toyota Research Institute, Keysight Technologies, Hitachi, Northrop Grumman, Amazon Web Services, Juniper Networks, NetApp, PG\&E, the Stanford Data Science Initiative, the Secure Internet of Things Project, and the NSF under grant EAR-1818579 and CAREER grant CNS-1651570.


\bibliographystyle{abbrv}
\Urlmuskip=0mu plus 1mu
\bibliography{main} 

\section*{APPENDIX}
\setcounter{section}{0}
 \setcounter{subsection}{0}
 \def\thesection{\Alph{section}}

\section{Comparison to Alternatives}
\label{appendix:alternate}

\subsection{Exact Similarity Search Algorithms}
\label{appendix:exact}
In this subsection, we investigate the performance and accuracy tradeoff between using MinHash LSH and exact algorithms for similarity search. We focus the comparison on set similarity joins, a line of exact join algorithms that identifies all pairs of sets above a similarity threshold from two collections of sets~\cite{setsimilarity}. State-of-the-art set similarity joins avoid exhaustively computing all pairs of set similarities via a filter-verification approach, such that only ``promising" candidates that survive the filtering and verification are examined for the final join.

We report single-core query time of our MinHash LSH implementation and four state-of-the-art algorithms for set similarity joins: PPJoin~\cite{ppjoin}, GroupJoin~\cite{groupjoin}, AllPairs~\cite{allpairs} and AdaptJoin~\cite{adaptjoin}. For the set similarity joins, we use an open-source implementation (C++) from a recent benchmark paper, which is reported to be faster than the original implementations on almost all data points tested~\cite{setsimilarity}.

We use a set of fingerprints generated from 20 hours of continuous time series data, which includes 74,795 input fingerprints with dimension 2048 and 10\% non-zero entries. For set similarity joins, we transform each binary fingerprint into a set of integer tokens of the non-zero entries, with the tokens chosen such that larger integer tokens are more frequent than smaller ones.  

We found that with a Jaccard similarity threshold of 0.5, the MinHash LSH incurs a 6.6\% false negative rate while enabling 63$\times$ to 200$\times$ speedups compared to set similarity join algorithms (Table~\ref{tab:setsim}). Among the four tested algorithms, AdaptJoin achieves the best query performance as a result of the small candidate set size enabled by its sophisticated filters. This is different from the benchmark paper's observation that expensive filters do not pay off and often lead to the slowest runtime~\cite{setsimilarity}. One important difference in our experiment is that the input fingerprints have a fixed number of non-zero entries; as a result, the corresponding input sets have equal length. Therefore, filtering and pruning techniques based on set length do not apply to our dataset. 

\subsection{Alternative LSH library}

\begin{table}
\small \center
\begin{tabular}{r r r r}
 \toprule
\textbf{False Negative (\%)} & \textbf{Query time (ms)} &\textbf{\# Hash Tables} & \textbf{\# Probes}  \\
\midrule
6.7 & 0.87 & 85 & 85    \\
 6.5 & 2.4 & 50 & 120  \\
0.54 & 2.4 & 50 & 400  \\
0.36 & 2.0 &  200 & 200 \\
\bottomrule
\end{tabular}
\caption{Average query time and false negative rate under different FALCONN parameter settings. }
\label{tab:falconn}
\end{table}

In this subsection, we compare the query performance of our similarity search to an alternative and more advanced open source LSH library. We were unable to find an existing high-performance implementation of LSH for Jaccard similarity, so we instead compare to FALCONN~\cite{falconnlib}, a popular library based on recent theoretical advances in LSH family for cosine similarity~\cite{falconn}.

We exclude hash table construction time, and compare single-core query time of FALCONN and our MinHash LSH. We use the cross-polytope LSH family and tune the FALCONN parameters such that the resulting false negative rate is similar to that of the MinHash LSH (6.6\%). With ``vanilla" LSH, FALCONN achieves an average query time of 0.87ms (85 hash tables); with multi-probe LSH, FALCONN achieves an average query time of 2.4ms (50 hash tables and 120 probes). In comparison, our implementation has an average query time of 36 $\mu$s (4 hash functions, 100 hash tables), which is 24$\times$ and 65$\times$ faster than FALCONN with vanilla and multi-probe LSH. We report the runtime and false negative rate under additional FALCONN parameter settings in Table~\ref{tab:falconn}. Notably, in multi-probe LSH, adding additional probes reduces the false negative rate with very little runtime overhead. We consider using multi-probe LSH to further reduce the memory usage as a valuable area of future work. 

The performance difference reflects a mismatch between our sparse, binary input and FALCONN's target similarity metrics in cosine distance. Our results corroborate previous findings that \mh outperforms SimHash on binary, sparse input data~\cite{minhashsimhash}.

\subsection{Supervised Methods}
\label{appendix:model}

In this subsection, we report results from using supervised models for earthquake detection on the Diablo Canyon dataset.

\minihead{Models} We focus the evaluation on two supervised models: WEASEL~\cite{weasel} and ConvNetQuake~\cite{convquake}. The former is a time series classification model that leverages statistics tests to select discriminative bag-of-pattern features on Fourier transforms; it outperforms the state-of-the-art non-ensemble classifiers in accuracy on the UCR time series benchmark. The latter is a convolutional neural network model with 8 strided convolution layers followed by a fully connected layer; it has successfully detected uncataloged earthquakes in Central Oklahoma.  

\minihead{Data} Same as the qualitative study in Section~\ref{sec:eq}, we focus on the area in the vicinity of the Diablo Canyon nuclear power plant in California. We use catalog earthquake events located in the region specified by Figure~\ref{fig:eqloc} as ground truth. We perform classification on the continuous ground motion data recorded at station PG.LMD, which has the largest number of high-quality recordings of catalog earthquake signals, and use additional data from station PG.DCD (station that remained active for the longest) for augmentation. Both stations record at 100Hz on 3 channels, capturing ground motion along three directions: EHZ channel for vertical, EHN channel for North-South and EHE channel for East-West motions. We use the vertical channel for WEASEL, and all three channels for ConvNetQuake. 

\minihead{Preprocessing and Augmentation} We extract 15-second long windows from the input data streams, which include windows containing earthquake events (positive examples) as well as windows containing only seismic noise (negative examples). This window length is consistent with that used for fingerprinting.

We adopt the recommended data preprocessing and augmentation procedures for the two models. For WEASEL, we z-normalize each 15-second window of time series by subtracting the mean and dividing by the standard deviation. For ConvNetQuake, we divide the input into monthly streams and preprocess each stream by subtracting the mean and dividing by the absolute peak amplitude; we generate additional earthquake training examples by perturbing existing ones with zero-mean Gaussian noise with a standard deviation of 1.2. For both models, we further augment the earthquake training set with examples of catalog events recorded at an additional station. 

In order to prevent the models from overfitting to location of the earthquake event in the time window (e.g. a spike in the center of the window indicates earthquakes), we generate 6 samples for each catalog earthquake event with the location of the earthquake event shifted across the window. Specifically, we divide the 15-second time window into five equal-length regions, and generate one training example from each catalog event with the event located at a random position within each region; we generate an additional example with earthquake event located right in the center of the window. We report prediction accuracy averaged on samples located in each of the five regions for each event. We further analyze the impact of this augmentation in the results section below. 

\minihead{Train/Test Split} We create earthquake (positive) examples from the arrival times from the Northern California Seismic Network (NCSN) catalog~\cite{NCEDC}. Together, the catalog yields 3585 and 1388 catalog events for PG.LMD and PG.DCD, respectively, from 2007 to 2017. We select a random 10\% of the catalog events from PG.LMD as the test set, which includes 306 events from 8 months. We create a second test set containing 449 new earthquake events detected by our pipeline. Both test sets exhibit similar magnitude distribution, with majority of the events centered around magnitude 1. The training set includes the remaining catalog events at PG.LMD, as well as additional catalog events at PG.DCD. 

For negative examples, we randomly sample windows of seismic noise located between two catalog events at station PG.LMD. For training, we select 28,067 windows of noise for WEASEL, and 874,896 windows for ConvNetQuake; ConvNetQuake requires a much larger training set to prevent overfitting. For testing, we select 85,060 windows of noise from September, 2016 for both models. 

Finally, we generate 15-second non-overlapping windows from one month of continuous data (December, 2011) in the test set. We then select 100 random windows that the model classifies as earthquakes for false positive evaluation. 

\minihead{Results} We report the two models' best classification accuracy on test noise events (true negative rate), catalog events and FAST events in Table~\ref{tab:model}. The additional training data from PG.DCD boosts the classification accuracy for catalog and FAST events by up to 4.3\% and 3.2\%. If the model is only trained on samples with the earthquake event in the center of the window, the accuracy further degrades for over 6\% for WEASEL and over 20\% for ConvNetQuake, indicating that the models are not robust to translation. 

Overall, the 20\% gap in prediction accuracy between catalog events and FAST events suggests that models trained on the former do not generalized as well to the latter. Since the two test sets have similar magnitude distributions, the difference indicates that FAST events might be sufficiently different from the existing catalog events in training set that they are not detected effectively. 

In addition, we report the false positive rate evaluated on a random sample of 100 windows predicted as earthquakes by each model. The ground truth is obtained via our domain collaborators' manual inspection. WEASEL and ConvNetQuake exhibit a false positive rate of 90\% with a 95\% confidence interval of 5.88\%. In comparison, our end-to-end pipeline has only 8\% false positives. 

\minihead{Discussion} The fact that unsupervised method like our pipeline is able to find qualitatively different events than those in the existing catalog suggests that, for the earthquake detection problem, supervised and unsupervised methods are not mutually exclusive, but complementary to each other. In areas with rich historical data, supervised models showed promising potential for earthquake classification~\cite{convquake}. However, in cases where there are not enough events in the area of interest for training, we can still obtain meaningful detections via domain-informed unsupervised methods. In addition, unsupervised methods can serve as a means for label generation to improve the performance of supervised methods.  

\section{Additional Evaluations}
\label{appendix:eval}
This section contains additional evaluation results for the factor analysis in Section 8.1, the microbenchmarks of pipeline parameters in Section 8.3 as well as a figure illustrating the key idea behind locality-sensitive hashing. 

In Table~\ref{tab:factor}, we report the runtime and relative improvement of each optimization in the factor analysis in Section 8.1 on 1 year of time series data at station LTZ in the New Zealand dataset.

\begin{table}[t]
\scriptsize
\begin{tabular}{l l l l l}
\toprule
    \textbf{Stages} & \textbf{Fingerprint} & \textbf{Hash Gen} & \textbf{Search} & \textbf{Alignment}   \\
    \midrule
    Baseline & 9.58 & 4.28 & 149 & $>$1 mo (est.) \\
    + occur filter & 9.58 & 4.28 & \textbf{30.9} (-79\%) & \textbf{16.02} \\
    + \#n func & 9.58 & \textbf{5.63} (+32\%) & \textbf{3.35} (-89\%) & \textbf{18.42} (+15\%)\\
    + locality Min-Max  & 9.58 & \textbf{1.58} (-72\%) & 3.35 & 18.42 \\
    + MAD sample & \textbf{4.98} (-48\%) & 1.58 & 3.35 & 18.42\\
    + parallel (n=12) & \textbf{0.54} (-89\%) & \textbf{0.14} (-91\%) & \textbf{0.62} (-81\%) & \textbf{2.25} (-88\%)\\
\bottomrule
\end{tabular}
\caption{Factor analysis (runtime in hours, and relative improvement) of each optimization on 1 year of data from station LTZ. Each optimization contributes meaningfully to the speedup of the pipeline, and together, the optimizations enable an over 100$\times$ end-to-end speedup. }
\label{tab:factor}
\end{table}

In Table~\ref{tab:madsample}, we report the relative speed up in MAD calculation time as well as the average overlap between the binary fingerprints generated using the sampled MAD and the original MAD as a metric for accuracy. The results illustrate that runtime reduces linearly with sampling rate, as expected. At lower rates, I/O begins to dominate MAD calculation runtime so the runtime improvements suffer from diminishing return. 
\begin{table}[t]
\small
\centering
\begin{tabular}{r r r}
\toprule
\textbf{Sampling Rate} & \textbf{Accuracy (\%)} & \textbf{Speedup}  \\
\midrule
0.001 & 94.9 & 350$\times$ \\
0.01 & 98.7 & 99.8$\times$ \\
0.1 & 99.5 & 10.5$\times$ \\
0.5 & 99.7 & 2.2$\times$\\
0.9 & 99.9 & 1.1$\times$\\
\bottomrule
\end{tabular}
\caption{Speedup and quality of different MAD sampling rate compared to no sampling on 1.3M fingerprints. Sampling enables a 100x speed up in MAD calculation with 98.7\% accuracy. Below 1\%, runtime improvements suffer from a diminishing return, as the IO begins to dominate the MAD calculation in runtime. }
\label{tab:madsample}
\end{table}

Finally, Figure~\ref{fig:lsh} illustrates the key difference between LSH and general hashing: LSH hash functions preserve the distance of items in the high dimensional space, such that similar items are mapped to the same ``bucket" with high probability. 

\section{Bandpass filter guidelines}
\label{appendix:bp}

\begin{figure}
\includegraphics[width=\linewidth]{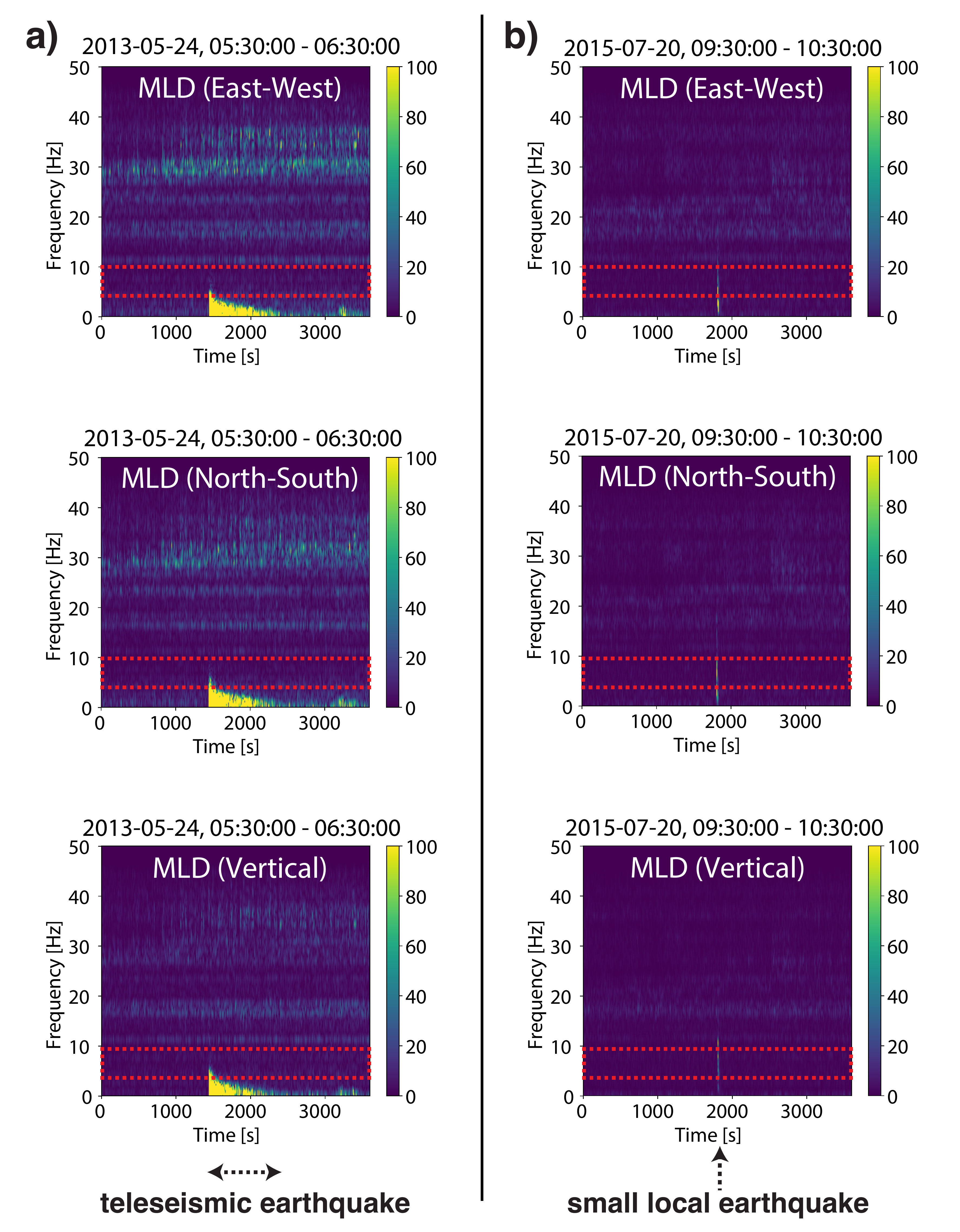}
\caption{Example hour-long spectrograms from the three components of continuous seismic data, sampled at 100 Hz, at station MLD from the Diablo Canyon, California, data set: East-West (top row), North-South (center row), vertical (bottom row).  For this station, a 4-10 Hz bandpass filter (dotted red rectangle) was applied before entering the processing pipeline. (a) Example of signals that should be excluded by the bandpass filter: a magnitude 8.3 teleseismic earthquake from the Sea of Okhotsk (bordered by Japan and Russia) starting at time ~1400 seconds, and persistent repeating noise throughout the entire hour at higher frequencies. (b) Example of a signal that should be included in the bandpass filter: a small local earthquake, with magnitude 1.7, at time ~1800 seconds.}
\label{fig:bp}
\end{figure}

Figure~\ref{fig:bp} illustrates the process of selecting the bandpass filter on an example data set. The provided examples are hour-long spectrograms computed from the three components of continuous seismic data at station MLD from the Diablo Canyon, California, data set. 

Figure~\ref{fig:bp}a shows examples of signals that should be excluded by the bandpass filter. The high-amplitude signal starting at time ~1400 seconds is from a magnitude 8.3 teleseismic earthquake near Japan and Russia, with a long duration of over 10 minutes and predominantly lower frequency content (below 4 Hz).  Generally, we are not interested in detecting large teleseismic earthquakes, because they are already detected and cataloged by global seismic networks (and shaking is usually felt near their origin).  There is also persistent repeating noise throughout the entire hour at higher frequencies: it is especially prominent at 30-40 Hz on the East-West and North-South channels, but there are several bands of repeating noise, starting at a low of 12 Hz.  We commonly observe repeating noise at lower frequencies (0-3 Hz) at most seismic stations, which is also seen in Figure~\ref{fig:bp}a after the teleseismic earthquake. It is essential to exclude as much of this persistent repeating noise from the bandpass filter as possible; otherwise, most of the fingerprints would match each other based on similar noise patterns, degrading both detection performance and runtime.

Figure~\ref{fig:bp}b shows an example of a small (magnitude 1.7) local earthquake signals, at time ~1800 seconds, that we would like to detect, and therefore should be included by the bandpass filter.  A small local earthquake is much shorter in duration, typically a few seconds long, and has higher frequency content, up to 10-20 Hz, compared to a teleseismic earthquake. We choose the widest possible bandpass filter to keep as much of the desired local earthquake signal as we can, while excluding frequencies with persistent repeating noise.

Figure~\ref{fig:bp}a and b show spectrograms from two different days (2013-05-24 and 2015-07-20) at one example seismic station. In general, we recommend randomly sampling and examining short spectrogram sections throughout the entire duration of available continuous seismic data, and at each seismic station used for detection, as the amplitudes and frequencies of the repeating noise can vary significantly over time and at different stations. Anthropogenic (cultural) noise levels are often higher during the day than at night, and higher during the workweek than on the weekend.  Sometimes it is difficult to select a frequency range that does not contain any persistent repeating noise; in this case, we advise excluding frequency bands with the highest amplitudes of repeating noise.

\section{Hash Signature Generation}
\label{appendix:hash}

We present pseudocode for the optimized hash signature generation procedure in Algorithm~\ref{alg:minmax}.

\begin{algorithm}[]
\footnotesize
\begin{algorithmic}
\Function{single\_hash}{d, t, k, seed} \Comment{Get all hash mappings}
\For {x $\in$ \{1, 2, ..., d\}} 
\For {y $\in$ \{1, 2, ..., t * k\}}
\State hash[i][j] = \textproc{murmurhash}(i, seed + j)
\EndFor
\EndFor
\Return hash
\EndFunction

\Function{minmax\_batch}{fp, hash} \Comment{Get hash signature for given batch}
\For {x $\in$ \{1, 2, ..., fp.size()\}} 
\For {y $\in$ \{1, 2, ..., d\}}
\If {fp[x][y] == 1} 
\For {i $\in$ \{1, 2, ..., t\}} 
\For {j $\in$ \{1, 2, ..., $\lceil\frac{k}{2}\rceil$ \} } 
\State minvals[i][j] = min(hash[y][i][j], minvals[i][j])
\State maxvals[i][j] = max(hash[y][i][j], maxvals[i][j])
\EndFor
\EndFor
\EndIf
\EndFor

\For {i $\in$ \{1, 2, ..., t\}} 
\State minmaxhash[x][i] = \textproc{hash\_combine}(minvals[i], maxvals[i])
\EndFor
\EndFor
\Return minmaxhash
\EndFunction
\\
\Function{gen\_signature}{fp, nprocs} \Comment{main function}
\State hash = \textproc{single\_hash}(d, t, $\frac{k}{2}$, seed)
\State fp\_partition = \textproc{partition}(fp, nprocs)
\For {i $\in$ \{1,2, ..., nprocs\}} \textbf{in parallel}
\State \textproc{minmax\_batch}(fp\_partition[i], hash)
\EndFor
\EndFunction

\end{algorithmic}
\caption{Optimized and parallelized Min-Max hash generation}
\label{alg:minmax}
\end{algorithm}

\end{document}